\documentclass[11pt,a4paper]{article}
\usepackage{jheppub}
\usepackage{amsfonts}
\usepackage[applemac]{inputenc}
\usepackage[mathscr]{eucal}
\usepackage{subfigure}
\usepackage{url}
\usepackage{ulem}
\usepackage{here}
\usepackage{mciteplus}
\usepackage{feynmp}
\usepackage{xspace}
\usepackage{slashed,color}

\newcommand{\sfd}[1]{{\bf #1}}
\newcommand{\sfb}[1]{{\bf {\bar #1}}}
\newcommand{\spa}[1]{\ensuremath{\widetilde{#1}}}
\newcommand\SARAH{{\tt SARAH}\xspace}

\newcommand\SPheno{{\tt SPheno}\xspace}
\newcommand\Vevacious{{\tt Vevacious}\xspace}

\newcommand\HB{{\tt HiggsBounds}\xspace}
\newcommand\HS{{\tt HiggsSignals}\xspace}
\newcommand\SSP{{\tt SSP}\xspace}

\newcommand\FlavorKit{{\tt FlavorKit}\xspace}
\newcommand{\sL}[0]{\ensuremath{{\tilde{\tau}}_{L}}\xspace}
\newcommand{\sE}[0]{\ensuremath{{\tilde{\tau}}_{R}}\xspace}
\newcommand{\sTL}[0]{\ensuremath{{\tilde{t}}_{L}}\xspace}
\newcommand{\sTR}[0]{\ensuremath{{\tilde{t}}_{R}}\xspace}
\newcommand{\sDRi}[0]{\ensuremath{{\tilde{d}^i}_{R}}\xspace}
\newcommand{\sDRj}[0]{\ensuremath{{\tilde{d}^j}_{R}}\xspace}
\newcommand{\sDLi}[0]{\ensuremath{{\tilde{d}^i}_{L}}\xspace}
\newcommand{\sDLj}[0]{\ensuremath{{\tilde{d}^j}_{L}}\xspace}

\newcommand{\vd}[1]{\ensuremath{v_{d}^{#1}}\xspace}
\newcommand{\vu}[1]{\ensuremath{v_{u}^{#1}}\xspace}
\newcommand{\vl}[1]{\ensuremath{v_{\tau_L}^{#1}}\xspace}
\newcommand{\ve}[1]{\ensuremath{v_{\tau_R}^{#1}}\xspace}

\newcommand{\vTL}[1]{\ensuremath{v_{t_L}^{#1}}\xspace}
\newcommand{\vTR}[1]{\ensuremath{v_{t_R}^{#1}}\xspace}
\newcommand{\vDLi}[1]{\ensuremath{v_{D^i_L}^{#1}}\xspace}
\newcommand{\vDRi}[1]{\ensuremath{v_{D^i_R}^{#1}}\xspace}
\newcommand{\vDLj}[1]{\ensuremath{v_{D^j_L}^{#1}}\xspace}
\newcommand{\vDRj}[1]{\ensuremath{v_{D^j_R}^{#1}}\xspace}

\newcommand{\lam}{\lambda}

\graphicspath{{figs/}}

\title{Resurrecting light stops after the 125 GeV Higgs in the baryon number violating CMSSM}

\author[a,b]{N.~Chamoun,}
\author[b]{H.~K.~Dreiner,}
\author[b]{F.~Staub}
\author[b]{and T.~Stefaniak}

\affiliation[a]{Physics Department, HIAST, P.O.Box 31983, Damascus,
Syria}
\affiliation[b]{Bethe Center for Theoretical Physics and Physikalisches Institut, Universit\"at Bonn, Bonn, Germany}

\emailAdd{nchamoun@th.physik.uni-bonn.de}
\emailAdd{dreiner@uni-bonn.de}
\emailAdd{fnstaub@th.physik.uni-bonn.de}
\emailAdd{tim@th.physik.uni-bonn.de}

\abstract{In order to accommodate the observed Higgs boson mass in the CMSSM, the stops must either be
  very heavy or the mixing in the stop sector must be very large. Lower stop masses, possibly more accessible 
  at the LHC, still give the correct Higgs mass only if the trilinear stop mixing parameter $|A_t|$ is in the 
  multi--TeV range. Recently it has been shown that such large stop mixing leads to an unstable 
  electroweak vacuum which spontaneously breaks charge or color. In this work we therefore go beyond the CMSSM and 
  investigate the effects of including baryon number violating operators $\lambda'' \bar{\bf U}
  \bar {\bf D}\bar {\bf D}$ on the stop and Higgs sectors. We find that for $\lambda'' \simeq {\mathcal{O}}(0.3)$ 
  light stop masses as low as 220~GeV are consistent with the observed Higgs mass as well as flavour 
  constraints while allowing for a stable vacuum. The light stop in this scenario is often the lightest 
  supersymmetric particle. We furthermore discuss the importance of the one--loop corrections involving $R$-parity 
  violating couplings for a valid prediction of the light stop masses. }


\preprint{BONN-TH-2014-10}

\begin{document}

\maketitle

\section{Introduction}
Run I of the the Large Hadron Collider (LHC) is complete. To date, there is no evidence for superpartner 
particles as predicted in supersymmetry (SUSY) \cite{Martin:1997ns} or experimental indications of any 
other physics beyond the Standard Model (SM).\footnote{See for example the talk given by 
O.~Buchm\"uller at the EPS 2013 conference in Stockholm \url{https://indico.cern.ch/event/218030/
session/28/contribution/869/material/slides/}.} The simplest SUSY scenarios like the constrained minimal 
supersymmetric standard model (CMSSM) \cite{Nilles:1983ge} are under pressure by the ongoing non--discovery, leading to the exclusion of large areas of parameter space 
\cite{Bechtle:2013mda,Craig:2013cxa,Bechtle:2012zk,Buchmueller:2012hv}. 
The observed Higgs boson mass of $m_h\approx 126$~GeV \cite{Aad:2012tfa,Chatrchyan:2012ufa} is within 
the previous predicted allowed range for supersymmetric models, including the CMSSM~\cite{Carena:2000dp}. 
In order to have at least part of the SUSY spectrum moderately light and accessible at the LHC, \textit{i.e.}~stop masses of
$m_{\tilde t_1}\lesssim 500$~GeV, it is necessary to maximise the mixing parameter in 
the stop sector, $X_t$ \cite{Heinemeyer:1998np}. However, it has been pointed out that large stop mixing in the 
(C)MSSM with rather light stop masses suffers from an unstable electroweak vacuum, such that charge or 
colour would be broken in a cosmologically short time 
\cite{Camargo-Molina:2013sta,Blinov:2013fta,Chowdhury:2013dka,Camargo-Molina:2014pwa}.  A stable 
electroweak vacuum together with the correct Higgs mass implies a lower limit on the stop mass of at least 
$800$~GeV. At the same time the stop should not be too heavy, in order to avoid the fine--tuning related to the 
hierarchy problem, see for example Ref.~\cite{Birkedal:2004xi}.

These conclusions are restricted to the (C)MSSM. More recently, non-minimal SUSY models have gained more 
attention. For instance, singlet extensions which give additional tree--level contributions to the Higgs mass, 
soften significantly the little hierarchy problem of the MSSM and can accommodate a much smaller stop mixing 
while obtaining the correct Higgs mass
\cite{Chang:2005ht,Hall:2011aa,Delgado:2012yd,Perelstein:2012qg,Kaminska:2013mya,Ross:2012nr,Ross:2011xv,Lu:2013cta,Kaminska:2014wia}. However, there are also non-minimal SUSY models with the MSSM 
particle content with appealing 
properties. It has been pointed out that the MSSM together with $R$-parity violation (RpV)
\cite{Hall:1983id,Barbier:2004ez,Dreiner:1997uz,Allanach:2003eb} can significantly weaken the collider mass limits 
\cite{Allanach:2012vj,Asano:2012gj,Franceschini:2012za,Evans:2012bf} and provide a rich phenomenology 
\cite{Dreiner:1991pe,Dreiner:2011wm,Dreiner:2012np,Dreiner:2012wm}. It is the purpose of this paper to 
extend the (C)MSSM to allow for the $R$-parity baryon-number violating operators $\lam''_{ijk}\sfb U_i \sfb D_j
\sfb D_k$ and in this framework to determine the allowed stop mass regions, which give (a) the correct Higgs 
mass, (b) a charge and colour stable vacuum, and (c) fulfil all experimental constraints from flavour observables.   We show 
that in this case  it is possible to have light stop masses of a few hundred GeV together with a Higgs mass 
consistent with the LHC observations, but without introducing charge and colour breaking (CCB) minima. 

The paper is organised as follows: in sec.~\ref{sec:model} we introduce the model under consideration. 
In sec.~\ref{sec:introHiggs} we explain the connection between the Higgs mass, light stops and the occurrence of charge 
and colour breaking minima in the baryon number violating CMSSM. In sec.~\ref{sec:results} we present our 
numerical results, before we conclude in sec.~\ref{sec:conclusion}. In the appendices we provide our benchmark 
points, sec.~\ref{app:benchmark}, the one--loop $R$pV corrections to the squark masses, sec.~\ref{app:stop}, 
as well as the minimum of the scalar potential, the vacuum, of the MSSM in the presence of $\lam''_{ijk}\sfb U_i 
\sfb D_j\sfb D_k$ operators, sec.~\ref{app:ScalarPotential}.

\section{The MSSM with Baryon Number Violation}
\label{sec:model}
$R$-parity  is a discrete $Z_2$ symmetry of the MSSM which is defined as 
\cite{Farrar:1978xj,Hall:1983id,Barbier:2004ez,Dreiner:1997uz,Allanach:2003eb}
\begin{equation}
  \label{eq:RParity}
  R_P = (-1)^{3(B-L)+2s} \,,
\end{equation}
where $s$ is the spin of the field and $B$, $L$ are its baryon respectively lepton number. We consider in the 
following the 
$R$-parity conserving superpotential of the MSSM
\begin{equation}
 W_R= Y^{i j}_{e}\,\sfd L_i \sfb E_j \sfd H_d 
   +Y^{i j}_{d}\, \sfd Q_i \sfb D_j \sfd H_d + 
   Y^{i j}_{u} \sfd Q_i \sfb U_j \sfd H_u +
 \mu\, \sfd H_u \sfd H_d \, ,
 \label{eq:superpot}
\end{equation}
and extend it only by the renormalizable baryon number violating operators 
\cite{Weinberg:1981wj,Sakai:1981pk}
\begin{equation}
  \label{eq:superpot$R$pV}
 W_{\slashed B} = \frac 12\lam_{ijk}^{\prime\prime} \sfb U_i \sfb D_j \sfb D_k,
\end{equation}
which also violate $R$-parity. Here $i,j,k=1,2,3$ are generation indices, while we suppressed $SU(3)$ colour 
and $SU(2)$ isospin indices.  In both of the previous equations $\sfd L_i,\, \sfb E_j,\, \sfd Q_i,\, \sfb U_i,$ $\sfb 
D_i$, $\sfd H_d$, $\sfd H_u$ denote the left chiral superfields of the MSSM in the standard notation 
\cite{Allanach:2003eb}.  We thus have for the total superpotential
\begin{equation}
W_{\mathrm{tot}}=W_R + W_{\slashed B}\,.
\label{superpot}
\end{equation}
This superpotential arises for example from the unique discrete gauge anomaly-free hexality $Z_6^R$. This is a 
discrete R-symmetry\footnote{For a discussion of R-symmetries see for example 
Refs.~\cite{Chamseddine:1995gb,Lee:2011dya}.} and is derived and discussed in Ref.~\cite{Dreiner:2012ae}. 
The low-energy $\mu$ term given in Eq.~(\ref{eq:superpot}) is generated dynamically \cite{Kim:1983dt,Giudice:1988yz}. 

For the superpotential in Eq.~(\ref{superpot}) the proton is stable, since lepton number is conserved, and the 
proton thus has no final state to decay to. However, heavier baryons can decay via double nucleon decay and virtual 
gluino or neutralino exchange \cite{Goity:1994dq}, if $\lam''$ couplings to light quarks 
are non--vanishing. However, we concentrate in the following exclusively on $R$pV couplings which involve the top 
quark. These are presently just bounded by perturbativity constraints \cite{Allanach:1999ic}, but could contribute 
at the one--loop level to flavour violating processes if the SUSY masses are not too heavy. For SUSY masses in 
the TeV range these effects are usually very small and do not provide better limits~\cite{Dreiner:2013jta}. 
The 
main reason is that it is usually only possible to constrain products of $\lam''$ couplings by flavour observables. 
However, we are 
going to consider in the following the case of only one non-vanishing $\lam''$ at the GUT scale. Other couplings 
get induced via the RGE running because of the quark flavour violation but those remain small.

The corresponding standard soft supersymmetry breaking terms for the scalar fields $\spa L,\spa E,\spa Q, \spa 
U,\spa D, H_d$, $H_u$ and the gauginos $\spa{B},\spa{W},\spa{g}$ read
\begin{eqnarray}
\nonumber -\mathscr{L}_{\text{SB},R} &=& m_{H_u}^2 |H_u|^2 + m_{H_d}^2
|H_d|^2+ \spa{Q}^\dagger m_{\spa{Q}}^2 \spa{Q} +
\spa{L}^\dagger m_{\spa{L}}^2 \spa{L} + \spa{D}^\dagger
m_{\spa{D}}^2 \spa{D} + \spa{U}^\dagger m_{\spa{U}}^2 
\spa{U} + \spa{E}^\dagger m_{\spa{E}}^2 \spa{E} \nonumber \\ && + \frac{1}{2}\left(M_1 \, \spa{B}
\spa{B} + M_2 \, \spa{W}_a \spa{W}^a + M_3 \, \spa{g}_\alpha
\spa{g}^\alpha + h.c.\right) \nonumber \\ && 
+ (\spa{Q}
T_u\spa{U}^\dagger H_u +  \spa{Q} T_d \spa{D}^\dagger H_d + 
\spa{L} T_e \spa{E}^\dagger H_d + B_\mu H_u H_d + \text{h.c.}) \thickspace 
\label{softterms}\\
 -\mathscr{L}_{\slashed B} &=& \frac 12 T_{\lam,{ijk}}^{\prime\prime} \spa U_i
 \spa D_j \spa D_k + \text{h.c.} \,. 
\end{eqnarray}
Here we have suppressed all generation indices, except in the last $R$pV term. The $m_{\tilde F}^2$ are 3$\times$3 matrices and 
denote the squared soft masses of the scalar components $\tilde F$ of the corresponding chiral superfields $F$. 
The $T_{u,d,e}$ are  3$\times$3 matrices of mass-dimension one. They are trilinear coupling constants of the scalar fields, and can be written in terms of the 
standard $A$-terms \cite{Nilles:1982dy} if no flavour violation is assumed, $T_{ii}^f= A_{i}^f Y_f^{ii}$, with $i=1,2,3$, and no 
summation over repeated indices, and $f=e,u,d$. Similarly, for the baryon number violating term we have $T^{''}_{\lam,ijk}=A^{''}_{ijk}
\lam^{''}_{ijk}$, again with no summation. 

Already the general, $R$-parity conserving MSSM with massless neutrinos has 105 parameters beyond 
those of the SM  \cite{Haber:1997if}. In the $R$--parity violating sector, as shown, there are 36 
additional parameters. Note that  $\lam_{ijk}^{\prime\prime}$ and $T_{\lam,{ijk}}^{\prime\prime}$ are 
anti-symmetric in the last two indices and can be complex. In order to significantly reduce the number 
of free parameters, we study a 
constrained model similar to the $R$--parity conserving CMSSM. As usual, we demand that all 
soft--breaking masses are universal at the grand unification (GUT) scale, $M_{\mathrm{GUT}}=
\mathcal{O}(10^{16}\,\mathrm{GeV})$. In addition, we treat the soft-breaking $R$pV couplings $T_{\lam,{ijk}}
^{\prime\prime}$ in the same way as the trilinear soft-breaking couplings of the MSSM, \textit{i.e.}~we assume that it is 
proportional to the corresponding superpotential term at $M_{\mathrm{GUT}}$, with a universal proportionality constant $A_0$. Thus, 
our boundary conditions at $M_{\mathrm{GUT}}$ are
\begin{eqnarray}
m_0^2 &\equiv& m_{H_d}^2 = m_{H_u}^2 \,, \hspace{1cm} {\bf 1} m_0^2 \equiv m_{\spa{Q}}^2 = 
m_{\spa{D}}^2 = m_{\spa{U}}^2 = m_{\spa{E}}^2 = m_{\spa{L}}^2 \\
M_{1/2} &\equiv& M_1 = M_2 = M_3 \\
T_{\lam}^{\prime\prime} &\equiv& A_0 \lam^{\prime \prime} \,, \hspace{2.6cm} T_i \equiv  A_0 Y_i 
\hspace{0.2cm} 
\text{with} \hspace{0.2cm} i=e,d,u \,.
\end{eqnarray}
The parameters $\mu$ and $B_\mu$ are fixed by the minimisation conditions of the vacuum ground state and we always set $
\mu>0$. We furthermore assume that all CP--violating phases vanish.

\section{Light stop, the Higgs mass and vacuum stability}
\label{sec:introHiggs}
\subsection{Dominant (s)top corrections to the Higgs mass}
\begin{figure}[tb]
\includegraphics[width=0.5\linewidth]{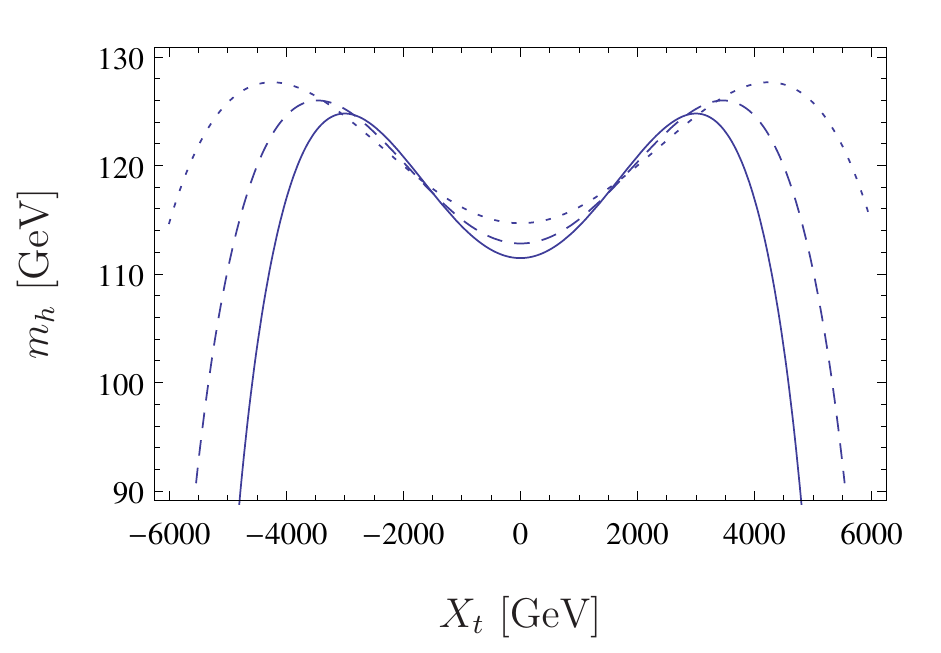} 
\hfill
\includegraphics[width=0.5\linewidth]{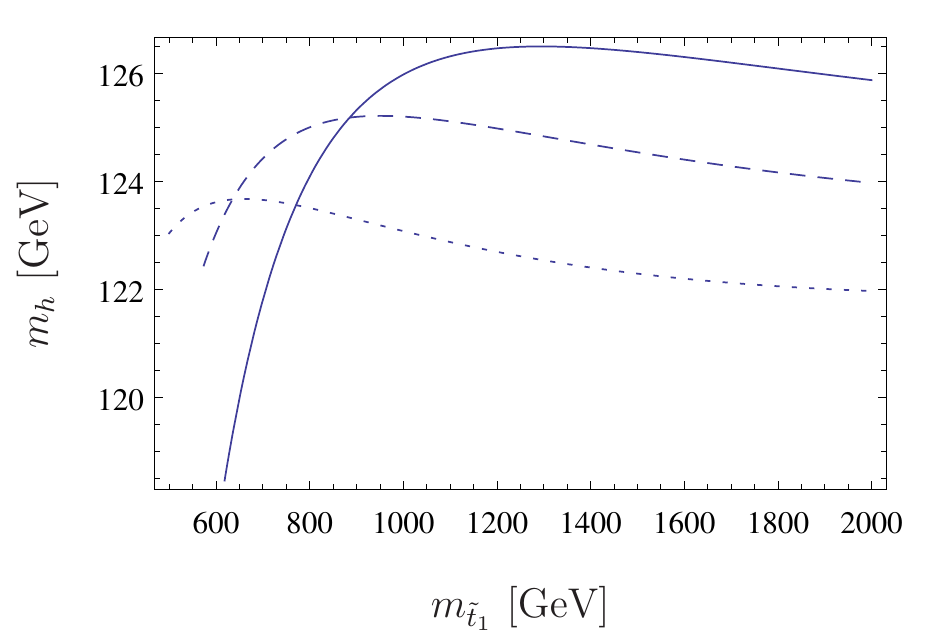} 
\caption{On the left: approximation of the light Higgs mass at one--loop as a function of $X_t$  with  
$m_{\tilde{t}_1}=750$~GeV (dotted),  $m_{\tilde{t}_1}=1000$~GeV (dashed),  $m_{\tilde{t}_1}=1500$~GeV (full). 
On the right: $m_h$ as a function of $m_{\tilde{t}_1}$  with  $X_t=-2.5$~TeV (dotted), $X_t=-3$~TeV (dashed), 
$X_t=-3.5$~TeV (full). In both plots we set $m_{\tilde{t}_2}=2$~TeV.} 
\label{fig:StopLoop}
\end{figure}

\begin{figure}
\includegraphics[width=0.5\linewidth]{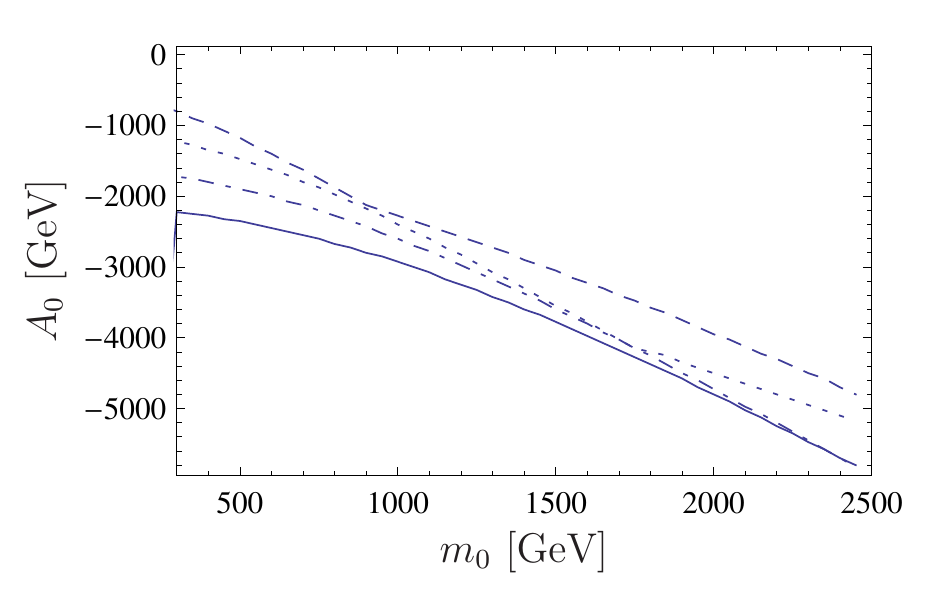} 
\hfill
\includegraphics[width=0.5\linewidth]{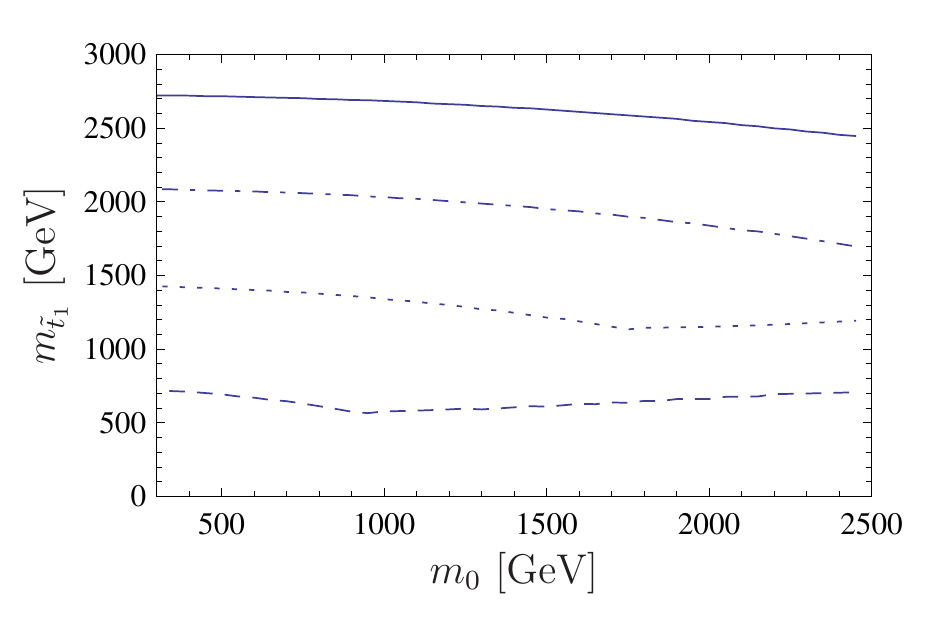} 
\caption{Left: the minimal value of $A_0$ in GeV compatible with a stable EW vacuum as a function of $m_0$ 
for $M_{1/2} = 0.5$~TeV (dashed), 1.0~TeV (dotted), 1.5~TeV (dot-dashed), 2.0 (full) and $\tan\beta=15$. Right: 
for the corresponding value of $A_0(m_0)$ of a given point in the left figure we compute the lightest 
stop mass. This is shown on the right as a function of $m_0$ for the same choices of $M_{1/2}$. We set $
\tan\beta=15$ in all cases.}
\label{fig:CMSSM}
\end{figure}

The main corrections to the light Higgs mass in the MSSM at one--loop stem from the (s)top contributions. They 
can be written in the decoupling limit $M_A \gg M_Z$ as 
\cite{Haber:1993an,Carena:1995wu,Martin:1997ns,Heinemeyer:1998np,Heinemeyer:1999be,Carena:2000dp}
\begin{equation}
\delta m_h^2 =\frac{3}{2 \pi^2} \frac{m_t^4}{v^2 }\left[\log \frac{M_S^2}{m_t^2} + \frac{X_t^2}{M_S^2} \left(1 - 
\frac{X_t^2}{12 M_S^2}\right) \right] 
\label{higgs-stop}
\end{equation}
with $M_S \equiv \sqrt{m_{\tilde{t}_1} m_{\tilde{t}_2}}$,  $m_t$ being the running $\overline{\text{DR}}$ top mass and 
$X_t \equiv A_t - \mu\cot\beta$. Our convention for the electroweak vacuum expectation value
(vev) is $v \simeq 246$~GeV. If one wants to 
keep $m_{\tilde{t}_i}$ moderately low (around or even below 1~TeV) the loop corrections required to explain the 
measured Higgs mass can be achieved by maximising the contributions proportional to the stop mixing $X_t$. 
$\delta m_h^2$ becomes maximal for $X_t = \sqrt{6} M_S$. In the following, we want to discuss the dependence 
of the light Higgs mass on $X_t$ and $m_{\tilde{t}_1}$ in more detail. For this purpose we show in 
Fig.~\ref{fig:StopLoop} the approximate values of the light Higgs mass at the one--loop level as function of these 
two parameters, respectively, keeping the other two fixed. The plots are based on the approximate formula given 
in Eq.~(\ref{higgs-stop}).
One can see that the light stop mass can be reduced by a few hundred GeV, for fixed values of $X_t$, without 
affecting the one--loop corrections to the light Higgs mass substantially. However, it is \textit{not} possible 
in the CMSSM to 
change $m_{\tilde{t}_1}$ without affecting $X_t$ and/or $m_{\tilde{t}_2}$, because all three parameter depend on 
$m_0$, $A_0$ and $\tan\beta$. This is problematic because it has recently been pointed out that the maximal 
mixing scenario, $X_t = \sqrt{6} M_S$, in the context of a light SUSY spectrum is ruled out by the instability of the 
electroweak (EW) vacuum:  The required large values of $|A_0|$ compared to $m_0$ lead to minima in the 
scalar potential below the EW vacuum, where colour and charge are broken by vacuum expectation values of 
stops or staus~\cite{Camargo-Molina:2013sta,Blinov:2013fta,Chowdhury:2013dka}. Furthermore, the EW  
vacuum would decay in a cosmologically short time. The condition of a stable EW vacuum can be used to put a 
lower limit on the light stop mass in the $R$-parity conserving CMSSM: one can determine the maximal value of $|
A_0|$ allowed by vacuum stability for fixed values of $\{m_0, M_{1/2}, \tan\beta\}$. This value can be translated 
into the minimal allowed stop mass for a given combination of $\{m_0, M_{1/2},\tan\beta\}$. These limits have 
been derived in Ref.~\cite{Camargo-Molina:2013sta} and we present examples in Fig.~\ref{fig:CMSSM}. For 
$m_0 >1$~TeV, $M_{1/2} > 1$~TeV it is not possible to get a light stop mass below 1~TeV. Lighter stops are 
possible for smaller values of $M_{1/2}$. However, this is often in conflict with current lower limits from gluino searches 
\cite{Aad:2014mra}. The constraint from the Higgs mass measurement has not yet been applied at this point.

\subsection{Vacuum stability and $R$-parity violation}
\begin{figure}
\includegraphics[width=0.5\linewidth]{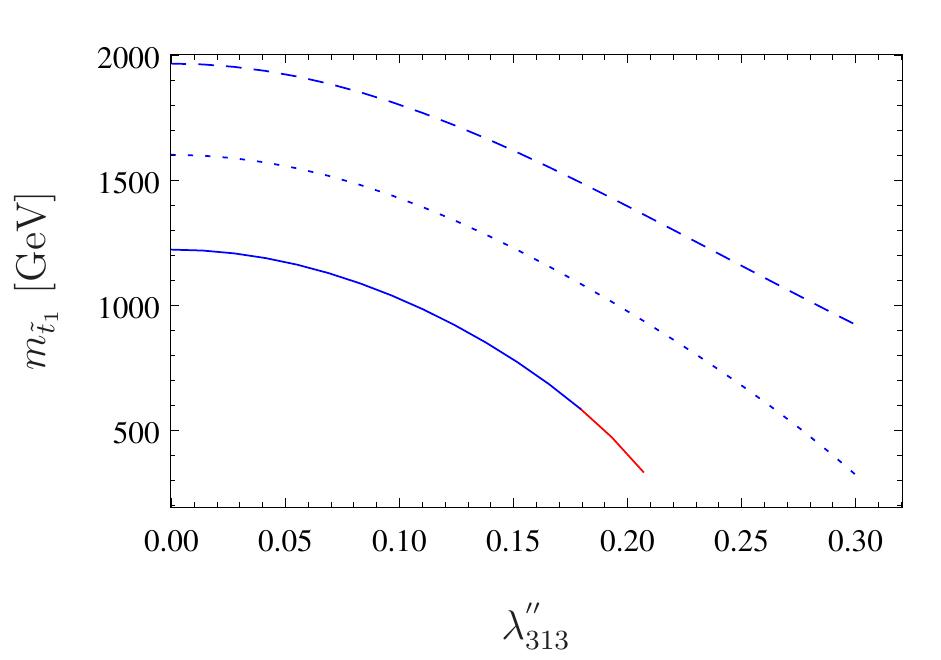} 
\hfill
\includegraphics[width=0.5\linewidth]{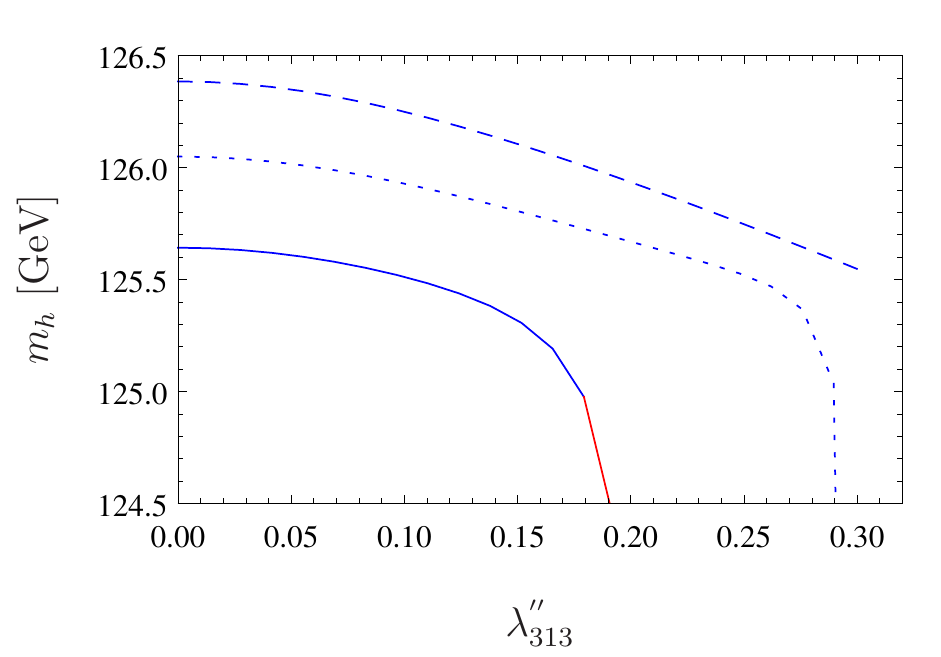}  \\
\includegraphics[width=0.5\linewidth]{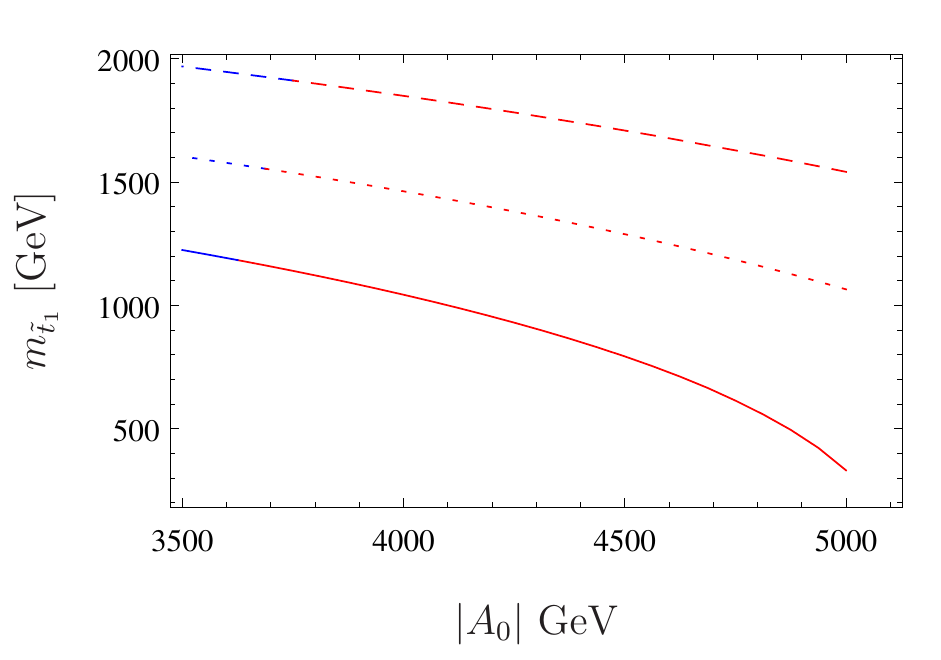} 
\hfill
\includegraphics[width=0.5\linewidth]{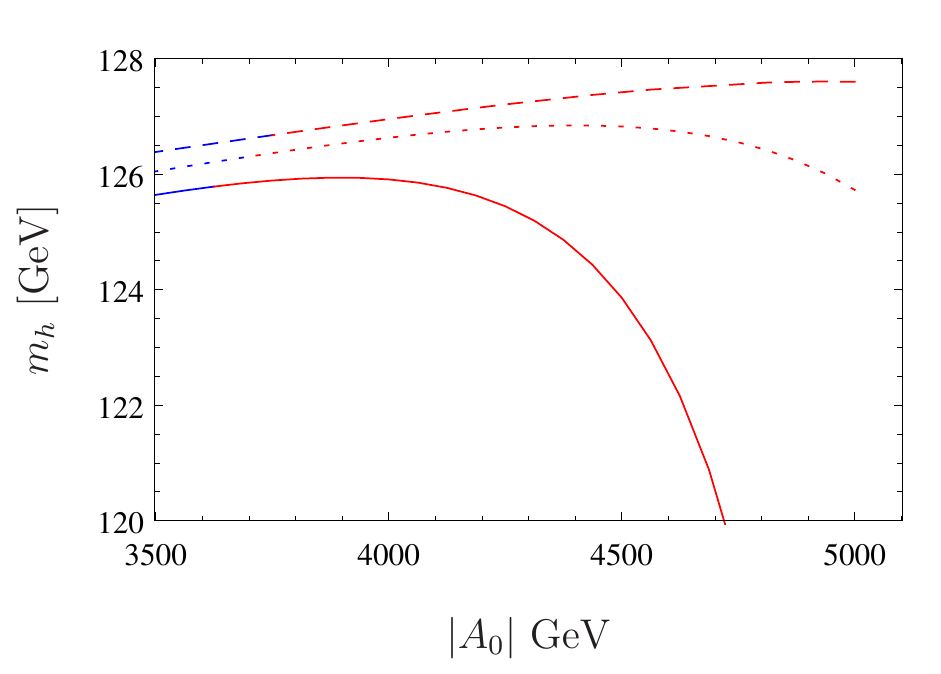} 
\caption{First row: the lightest stop mass, $m_{\tilde t_1}$, (left) and light Higgs mass, $m_h$, (right) as a function of the $R$-parity violating 
coupling $\lam^{''}_{313}$ evaluated at $M_{\mathrm{GUT}}$. Here, we set $m_0 = 1500$~GeV, $\tan\beta = 12$, 
$A_0 = -3500$~GeV and $\mu>0$. The solid lines are for $M_{1/2} = 1000$~GeV, the dotted lines are 
for $M_{1/2} = 1250$~GeV, and the dashed lines are for $M_{1/2} = 1500$~GeV. 
Second row: Dependence of $m_{\tilde t_1}$ (left) and $m_h$ (right) on $A_0$ 
(we consider only $A_0 < 0$ here) in the $R$-parity conserving case, $\lam_{313}^{''} = 0$. The remaining parameters are chosen as in the first row. The blue lines correspond to a stable and the red lines to a meta-stable EW vacuum.}
\label{fig:RunningStop}
\end{figure}

The situation described above changes if one allows for non-vanishing $R$-parity violating couplings. These 
affect at one--loop for example the running of $T_{u,33}=A_t Y_u^{tt}$ and $m_{u,33}^2 = m^2_{\tilde t_R}$~\cite{Allanach:2003eb}:
\begin{align}
\beta_{T_{u,ij}}^{(1)} & = \beta_{T_{u,ij}}^{(1),\text{MSSM}}+  2 
\lam^{''*}_{{a b c}} Y_{u,{a {j}}}  T^{''}_{\lam_{i b c}}  
+\lam^{''*}_{{c a b}} 
\lam^{''}_{{{i} a b}}  T_{u,{c {j}}}, \\
\label{eq:RGEmu2}
\beta_{m_{\tilde U,i j}^2}^{(1)} & = \beta_{m_{u,i j}^2}^{(1),\text{MSSM}} +2 T^{''*}_{\lam_{j a b}} 
T''_{\lam_{i a b}}  \nonumber \\ 
&\hspace{1cm} + 4 
\lam^{''*}_{{{j} a c}} \lam^{''}_{{{i} a b}} m_{\tilde D,{c b}}^{2}  + 
\lam^{''*}_{{c a b}} \lam^{''}_{{{i} a b}} m_{\tilde U,{c {j}}}^{2}  + 
\lam^{''*}_{{{j} b c}} \lam^{''}_{{a b c}} m_{\tilde U,{{i} a}}^{2}  .
\end{align}
To demonstrate the consequences of these additional terms, we show as an example the affect of 
$\lam^{''}_{313} (M_{\mathrm{GUT}})$ on the weak scale values of $m_{\tilde t_1}$ and $m_h$ in 
Fig.~\ref{fig:RunningStop}. We already distinguish here between points with a stable and an unstable EW 
vacuum. For these plots we have used our full numerical setup, explained in detail below in sec.~\ref{sec:setup}. 

Here, we start with a fixed set of CMSSM parameters $\{m_0,M_{1/2},\tan\beta,A_0\}$ with a stable EW vacuum 
and then turn on $\lam^{''}_{313}$. We see that the light stop mass can be reduced by several hundred GeV 
without spoiling the vacuum stability or affecting the light Higgs mass too much. For comparison, we show also 
the impact of a variation of $A_0$ while keeping $\lam_{313}^{''}=0$. This reduces also the light stop mass 
as expected and has a much larger impact on the Higgs mass. However, values of $A_0$ below --3.7~TeV are 
forbidden because the vacua where charge and colour are broken become deeper than the EW vacuum. 
Thus, baryon number violating couplings are a very attractive possibility to obtain
light stop scenarios in the CMSSM which are not in conflict with vacuum stability.

\subsection{Loop corrections to the stop mass}
\label{sec:stop}
We have seen that light stops and the correct Higgs mass can be obtained for large values of $\lam^{''}_{31i}$ if 
the operator couples directly to the top quark. However, the $R$-parity violating coupling will not only change the RGE 
running, as discussed in Ref.~\cite{Allanach:1999mh,Allanach:2003eb}, but also contribute to the radiative corrections 
to the stop masses at the one--loop level. Since these corrections to the stop masses  to our knowledge have not been considered so far in the literature, we discuss the effect here. The analytical calculation is summarised in 
Appendix~\ref{app:stop}. We find that the corrections to the right--stop mass squared are approximately given by
\begin{equation}
\delta m_{\tilde t_R}^2 \simeq \frac{1}{8\pi^2}  |\lam^{''}_{3ij}|^2 M_{SUSY}^2\,.
\label{1loopmassrpv}
 \end{equation}
Here, $M_{SUSY}$ is taken to be the mass scale of the down--like squarks running in the loop. If these 
masses are much heavier than the stop they can give large positive contributions to the light stop mass.

\begin{figure}[tb]
\begin{picture}(0,0)
\put(35,30){\includegraphics[width=0.23\linewidth]{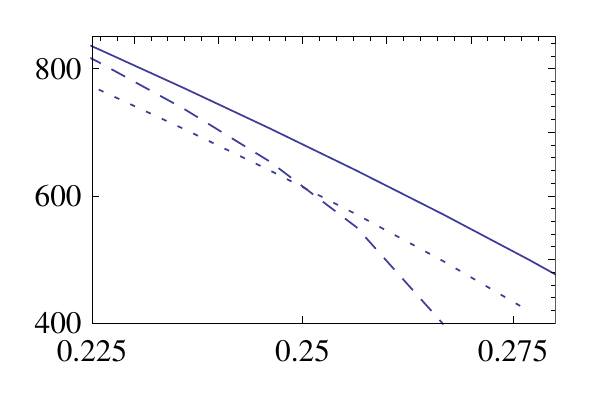}}
\end{picture}
\includegraphics[width=0.5\linewidth]{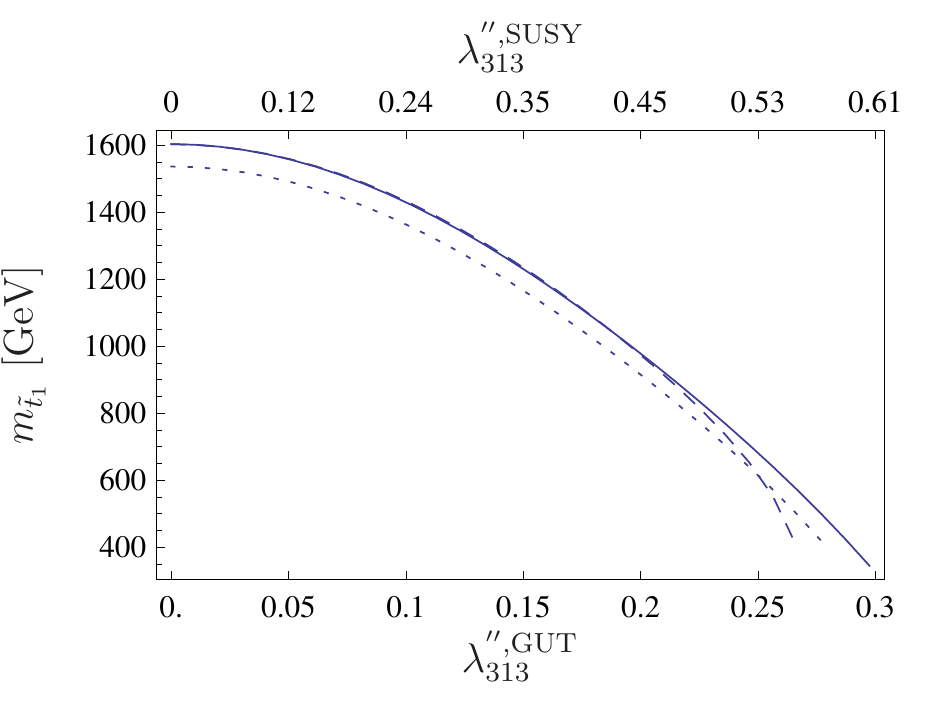} 
\hfill
\includegraphics[width=0.5\linewidth]{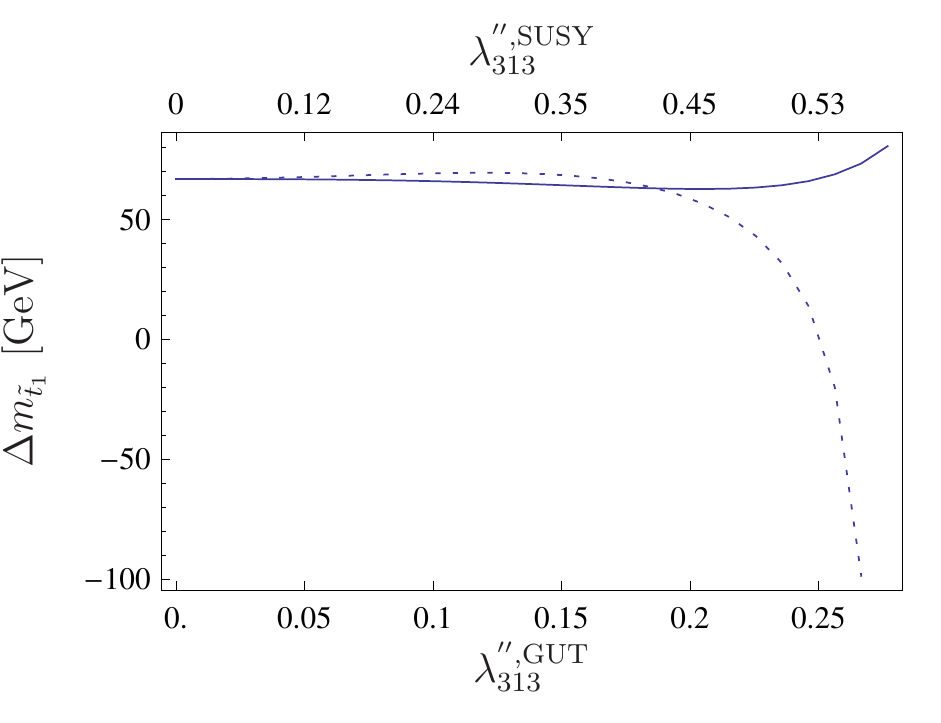}  \\
\includegraphics[width=0.5\linewidth]{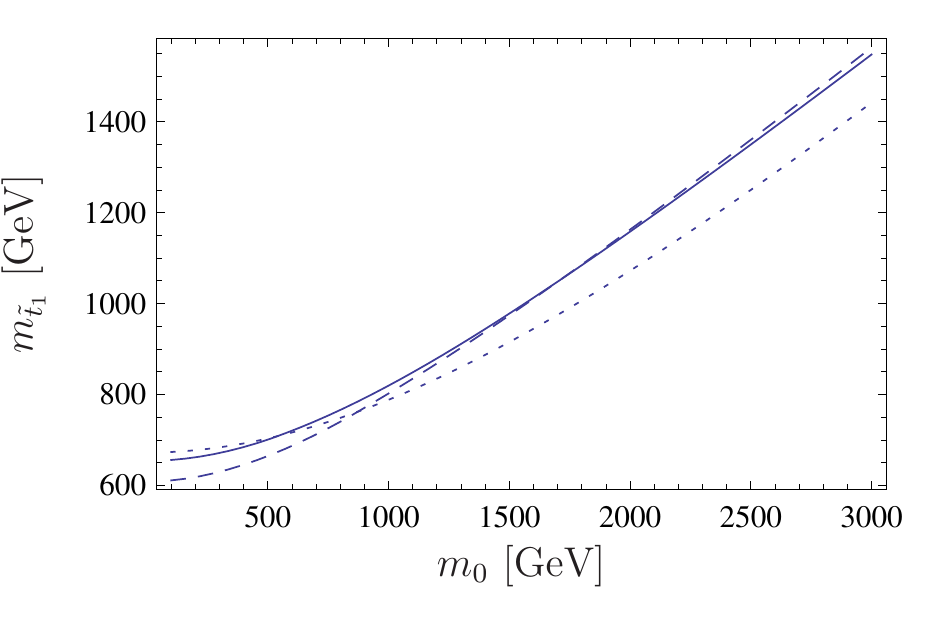} 
\hfill
\includegraphics[width=0.5\linewidth]{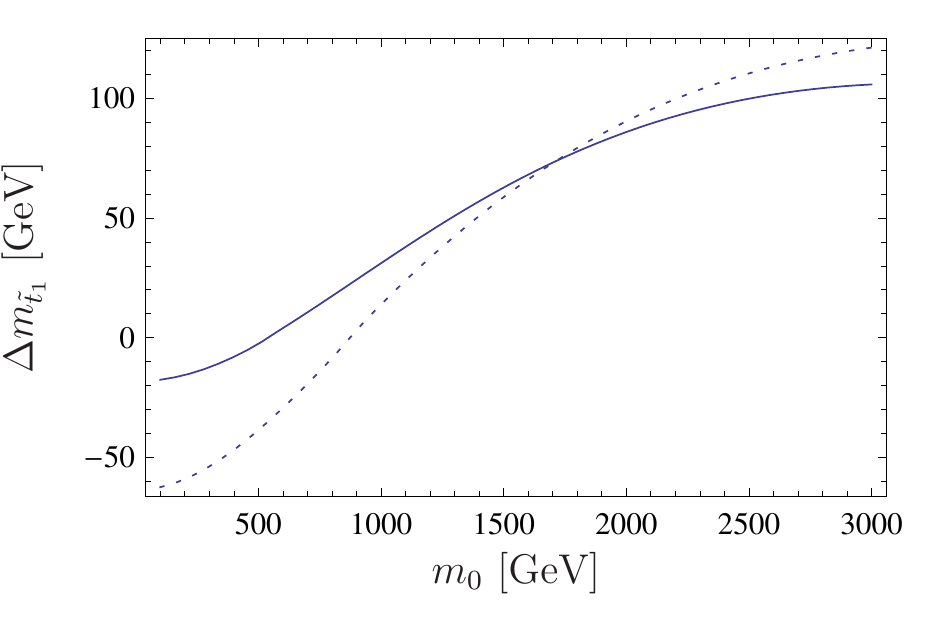} 
\caption{First row: the lightest stop mass, $m_{\tilde t_1}$, as a function of the $R$-parity violating
coupling $\lam^{''}_{313}$, evaluated either at $M_{\mathrm{GUT}}$ (bottom axis labels) or at 
$M_{\mathrm{SUSY}}$ (top axis labels). We set $m_0 = 1500$~GeV, $M_{1/2} = 1250$~GeV, $\tan\beta = 12$, 
$A_0 = -3500$~GeV, and $\mu>0$. On the left we show the tree--level mass (dotted line), the 
one--loop mass without $\bar{\bf U}\bar {\bf D}\bar {\bf D}$ corrections (dashed line) and the mass with full 
one--loop corrections (full line). On the right we show the mass difference between the tree--level and the 
one--loop mass ($\Delta m_{\tilde t_1} = m^{(1L)}_{\tilde t_1} -m^{(T)}_{\tilde t_1}$) with (full line) and without 
(dotted line) $\bar{\bf U}\bar {\bf D}\bar {\bf D}$ corrections. Note that the contributions of the $\bar{\bf U}\bar 
{\bf D}\bar {\bf D}$ operators to the RGE running are included in all cases. Second row: $m_{\tilde t_1}$ as 
function of $m_0$. Here we fixed $\lam_{313}^{'',\scriptsize\mbox{GUT}} = 0.2$. The other parameters are chosen 
as in the first row.}
\label{fig:Stop1Loop}
\end{figure}

To show the importance of these corrections we present in Fig.~\ref{fig:Stop1Loop} the mass of the light stop as 
function of $\lam^{''}_{313}$ and $m_0$ at tree and one--loop level.  We present in the top figures
$\lam^{''}_{313}$ evaluated at both $M_{\mathrm{GUT}}$ and $M_{\mathrm{SUSY}}$. At one--loop we give 
the results with and without the $R$-parity violating corrections to the stop self--energies. These results are 
based on a full numerical calculation which does not rely on the simplifying assumptions made in  
Appendix~\ref{app:stop}. All effects of flavour mixing, mass difference between squarks, and of the external 
momentum  are taken into account. The numerical calculation is based on the general procedure to calculate 
one--loop mass spectra with the {\tt Mathematica} package \SARAH, presented in Ref.~\cite{Staub:2010jh,Staub:2010ty}. 

We can see that for light stops the loop corrections are very important. They are dominated by the $\alpha_s$ 
corrections if the $\bar{\bf U}\bar {\bf D} \bar {\bf D}$ contributions are neglected. These corrections are negative 
and quickly reduce the tree--level mass in the limit $m_{\tilde g} \gg m_{\tilde t_1}$ \cite{Donini:1995wh}. In 
contrast, the $\bar{\bf U}\bar {\bf D}\bar {\bf D}$ corrections are positive and stabilise the stop mass at the 
one--loop level. We see that these corrections can easily shift the stop mass by more than 100~GeV compared 
to the case of an $R$-parity conserving one--loop calculation. Thus, these corrections have to be taken into 
account for any meaningful prediction of light stop masses in the context of large $R$-parity violating couplings.

\section{Results}
\label{sec:results}
We are interested in parameter regions in the CMSSM extended by baryon number violating operators, which 
provide a light stop. As constraints, we take the Higgs mass measurement, the limits on new physics from 
flavour observables and the stability of the EW vacuum. Before we present the preferred regions, we give the 
main details of our numerical setup. 

\subsection{Numerical setup}
\label{sec:setup}
We have used \SARAH  \cite{Staub:2008uz,Staub:2009bi,Staub:2010jh,Staub:2012pb,Staub:2013tta} 
to obtain a \SPheno \cite{Porod:2003um,Porod:2011nf} version of the MSSM with trilinear $R$pV. This \SPheno 
version calculates the renormalisation group equations (RGEs) taking into account the full $R$pV effects at the 
one-- and two--loop level. The RGEs have been cross checked against 
Refs.~\cite{Allanach:1999mh,Allanach:2003eb}. In addition, the program calculates the entire mass spectrum at 
the one--loop level. Thus, also the one--loop corrections to the stop masses  stemming from $\lam^{''}_{31i}$ 
discussed in sec.~\ref{sec:stop} are included. Furthermore, the known two--loop corrections to the light 
Higgs mass are taken into account \cite{Brignole:2001jy,Brignole:2002bz,Dedes:2002dy,Dedes:2003km}.  We 
have compared the mass spectrum calculation of this \SPheno version with {\tt SoftSusy 3.3.8} 
\cite{Allanach:2001kg,Allanach:2009bv}. We found good agreement if we remove the radiative $R$pV 
corrections to the mass spectrum, which are not included in {\tt SoftSusy}. The remaining differences are of the 
same size as the known discrepancies in the MSSM \cite{Allanach:2003jw} and provide an 
estimate of the theoretical uncertainty.

Moreover, \SPheno calculates the decay widths and branching ratios of the Higgs boson(s), as well as the Higgs 
couplings normalised to the SM expectations. We employ this information through \HB~\cite{Bechtle:2008jh,Bechtle:2011sb,Bechtle:2013gu,Bechtle:2013wla} and \HS~\cite{Bechtle:2013xfa,Bechtle:2014ewa} 
 to confront the Higgs sector for a given parameter point 
with existing measurements and exclusion limits. This has almost no influence on our results, as the stops we 
obtain are too heavy \cite{Belyaev:2013rza}.

There are also a wide range of flavour observables calculated by \SPheno with a high precision even for SUSY 
models beyond the MSSM, thanks to the \FlavorKit interface~\cite{Porod:2014xia},
which is an automatisation of 
the approach presented in Ref.~\cite{Dreiner:2012dh}. We consider in the following the observables $B \to X_s \gamma$, 
$B^0_q \to \mu^+ \mu^-$, and $\Delta M_{B_q}$ ($q=s,d$), which provide the best limits. To accept or discard 
parameter points based on the flavour observables we consider the ratio $R(X)$ defined as
\begin{equation}
 R(X) \equiv \frac{X}{X_{SM}}\,.
\end{equation}
Here $X$ is the predicted value of each flavour observable for a given parameter point, and $X_{SM}$ is the 
corresponding SM theoretical expectation. If we assume a 10\% uncertainty in the SUSY calculation and 
combining the experimental limits together with the corresponding SM predictions, we get the following 
constraints at 95\% C.L. on the $R(X)$:
\begin{itemize}
 \item $B \to X_s \gamma$ \cite{Beringer:1900zz,Misiak:2009nr,Asner:2010qj,Nakamura:2010zzi,Haisch:2012re}
 \begin{equation}
 0.89 < R(\text{BR}(b\to s \gamma)) < 1.33
 \end{equation}
 \item $B_q \to l^+ l^-$ \cite{Haisch:2012re,Chatrchyan:2013bka,Aaij:2013aka,Dreiner:2013jta}
  \begin{eqnarray}
  0.43  <& R(\text{BR}(B_s \to \mu^+ \mu^-)) & < 1.35 \\
   & R(\text{BR}(B_d \to \mu^+ \mu^-)) & <  8.3 
 \end{eqnarray}
 \item $\Delta M_{B_q}$ \cite{Buras:1990fn,Golowich:2011cx,Nakamura:2010zzi,Arana-Catania:2013pia}
   \begin{eqnarray}
  0.54 < & R(\Delta M_{B_s}) & < 1.44 \\
  0.25 < & R(\Delta M_{B_d}) & <  1.84
 \end{eqnarray}
\end{itemize}

To check the vacuum stability of each parameter point we use the computer program
\Vevacious \cite{Camargo-Molina:2013qva}, for 
which we have created the corresponding model files with \SARAH. For this step, we had to restrict 
ourselves to a set of particles, which can in principle get a non--vanishing vev. Since the required computational effort grows 
quickly with the number of allowed vevs, we employ a two step approach: First, we assume that only the 
staus and stops can have non-vanishing vevs (\vl{},\ve{}, \vTL{}, \vTR{}), besides those for the neutral Higgs 
scalar fields. All points which pass this check, \textit{i.e.}~they do not have a charge or colour breaking minimum, 
are again checked for the global vacuum taking into account the vevs of those generations of 
down--squarks which are involved in the $\lam^{''}_{3ij}$ coupling  (\vTL{}, \vTR{}, \vDLi{}, \vDRi{}, 
\vDLj{},\vDRj{}). Here it is necessary to allow also for vevs of the left--handed counterparts of the 
right down--like squarks to find $D$-flat directions in the scalar potential, even though they do not couple 
directly to $\bar{\bf U}\bar {\bf D}\bar {\bf D}$. The scalar potential at tree--level is discussed in 
more detail in Appendix~\ref{app:ScalarPotential}. As discussed there, the additional checks 
for CCB vacua with down--like squark vevs should have only a minor impact on the number of points which are 
ruled out by the vacuum considerations. This is confirmed by our numerical study: only 5\% of the points with 
stop masses below 1~TeV, which seem to be stable when checking only for stau and stop vevs, are in fact only 
meta--stable when including the sdown and sbottom vevs in addition ($\lam_{313}^{''}$--case). For  
$\lam_{312}^{''}$ no points are affected by the additional check for vacua with sdown and sstrange 
vevs. 

In the following, we only accept points which do not exhibit a deeper CCB vacuum. 
In principle, it might be possible that the EW vacuum is only meta--stable, but long--lived on a cosmological time 
scale. However, it is been shown that vacua which seem to be long--lived at zero temperature are likely to have 
tunnelled in the early universe into the CCB vacuum if temperature effects are taken into account 
\cite{Camargo-Molina:2014pwa}.

\subsection{Light stops in the CMSSM with $\bar{\bf U}\bar {\bf D} \bar {\bf D}$ operators}
\begin{figure}[tb]
\centering
\includegraphics[width=0.49\linewidth]{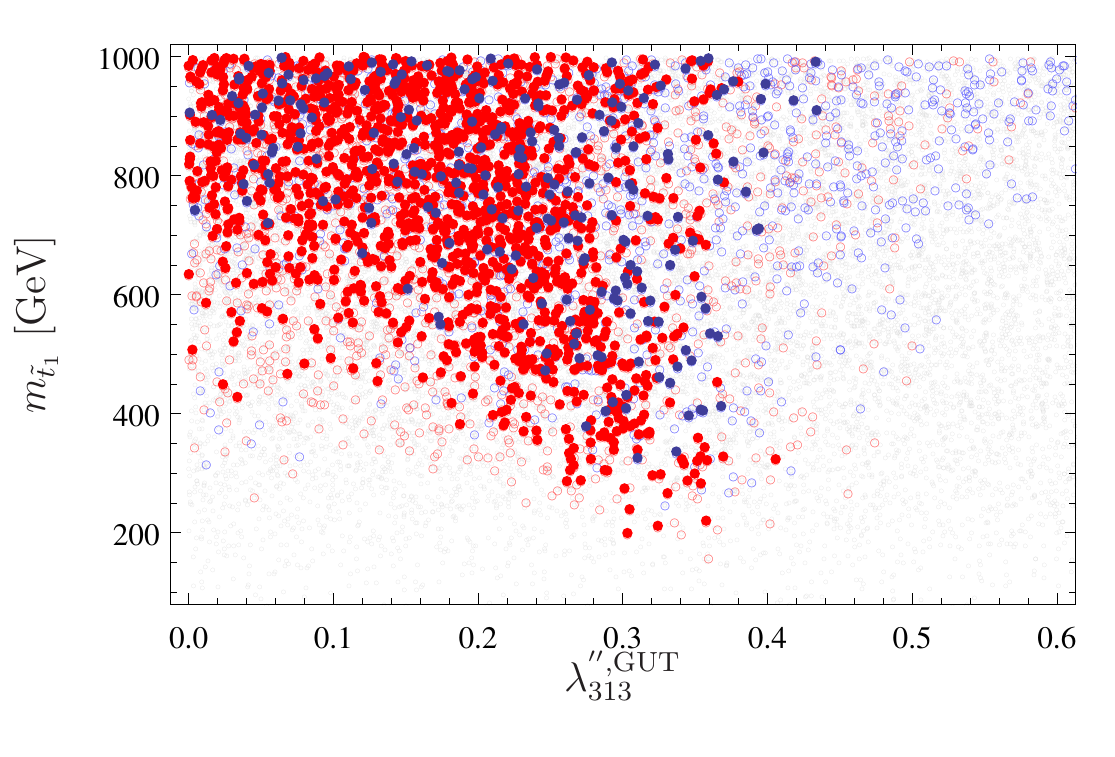}  \hfill 
\includegraphics[width=0.49\linewidth]{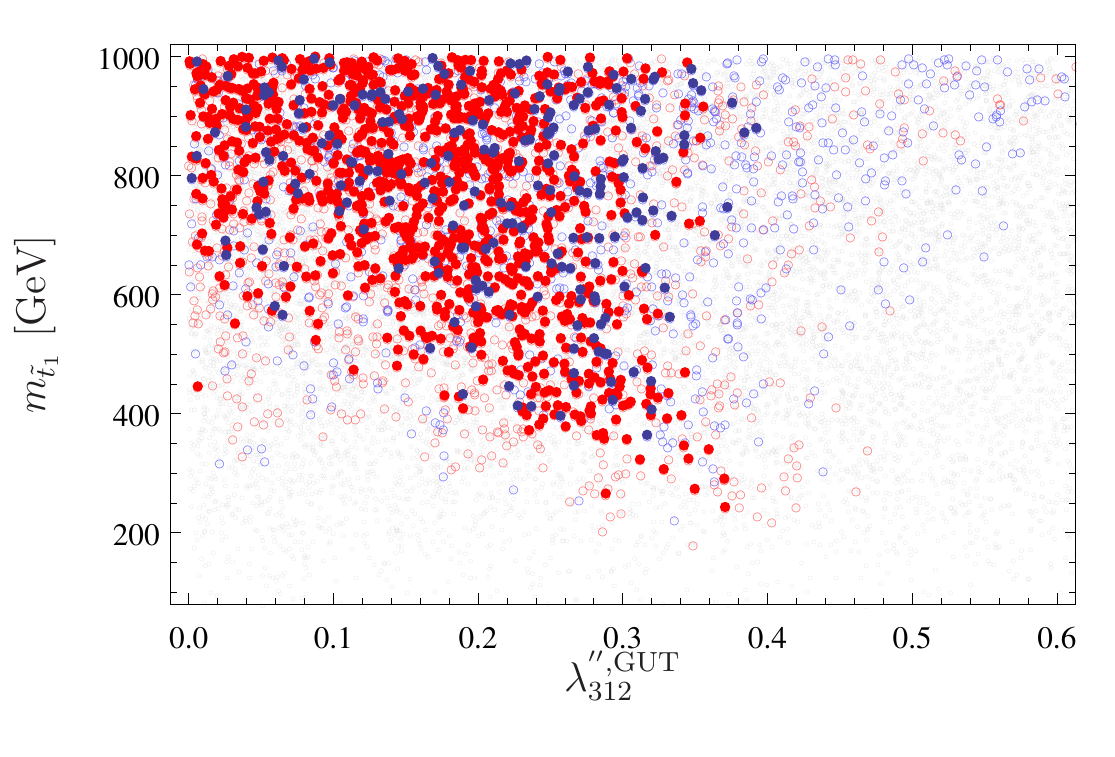} 
\caption{Vacuum stability in the $(\lam^{'',{\scriptsize\mbox{GUT}}}_{313},m_{\tilde{t}_1})$ (left) and $
(\lam^{'',{\scriptsize\mbox{GUT}}}_{312},m_{\tilde{t}_1})$ (right) planes based on a random scan using the 
parameter ranges of Eqs.~(\ref{eq:ranges}) and (\ref{eq:ranges2}). The blue points have a stable EW vacuum 
while for the red points deeper CCB vacua exist. For the filled dots we required  $m_h \in [124,128]$~GeV, and 
for the large empty circles we applied $m_h \in [122,130]$~GeV as a cut. The small empty grey circles are 
without any cut on the Higgs mass but have a stable vacuum.}
\label{fig:result313}
\end{figure}

To find regions with light stops in the CMSSM in the presence of $\bar{\bf U}\bar {\bf D}\bar {\bf D}$ operators in 
agreement with all constraints, we performed random scans with the tool \SSP 
\cite{Staub:2011dp} restricted to the following ranges of CMSSM parameters
\begin{equation}
m_0 \in [0,2]~\text{TeV,} \, \hspace{0.4cm}
M_{1/2} \in [0,2]~\text{TeV,} \, \hspace{0.4cm}
\tan\beta \in [5,60], \, \hspace{0.4cm}
A_0 \in [-10,0]~\text{TeV,} \, \hspace{0.4cm}
\mu > 0\,.
\label{eq:ranges}
\end{equation}
For the $\bar{\bf U}\bar {\bf D} \bar {\bf D}$ parameters we have chosen the range
\begin{equation}
\label{eq:ranges2}
\lam^{''}_{31i} \in [0,0.7] \hspace{2cm} \text{with} \, \, i=2,3\,,
\end{equation}
as the input at the GUT scale. The results are summarised in Fig.~\ref{fig:result313}. Here, we have 
applied two different cuts on the Higgs mass: $m_h = (126 \pm 4)$~GeV or the stricter case $m_h = (126 \pm 
2)$~GeV. The second cut is motivated by the theoretical uncertainty for the light Higgs mass of 2--3~GeV, which 
is usually assumed when using the known two--loop results. However, the $\lam''_{3ij}$ couplings give rise to 
new corrections to the Higgs mass at two--loop. These contributions are unknown and increase therefore the 
theoretical uncertainty. 

In Fig.~\ref{fig:result313}, the blue points have a stable EW vacuum while for the red points deeper CCB vacua 
exist. For the filled dots we required  $m_h \in [124,128]$~GeV, while for the large empty (red and 
blue) circles we applied the weaker constraint $m_h \in [122,130]$~GeV. The small empty grey circles are 
without any cut on the Higgs mass, but have a stable vacuum. The figure on the left differs from that on the 
right slightly, due to the lighter sbottom mass to be applied in Eq.~(\ref{1loopmassrpv}). The corresponding plot 
for $\lam^{''}_{323}$ is indistinguishable from that presented here for $\lam^{''}_{313}$. 

We can see the general trend: For increasing  $\lam^{''}_{31i}$ a lighter and lighter stop mass is compatible 
with all constraints. One central result of this paper is that we can have a stop mass as low as 220 GeV while satisfying 
the strict Higgs mass constraint and also having a stable EW vacuum, for $\lam^{''}_{31j}\gtrsim0.3$.

\begin{figure}[tb]
\centering
\includegraphics[width=0.49\linewidth]{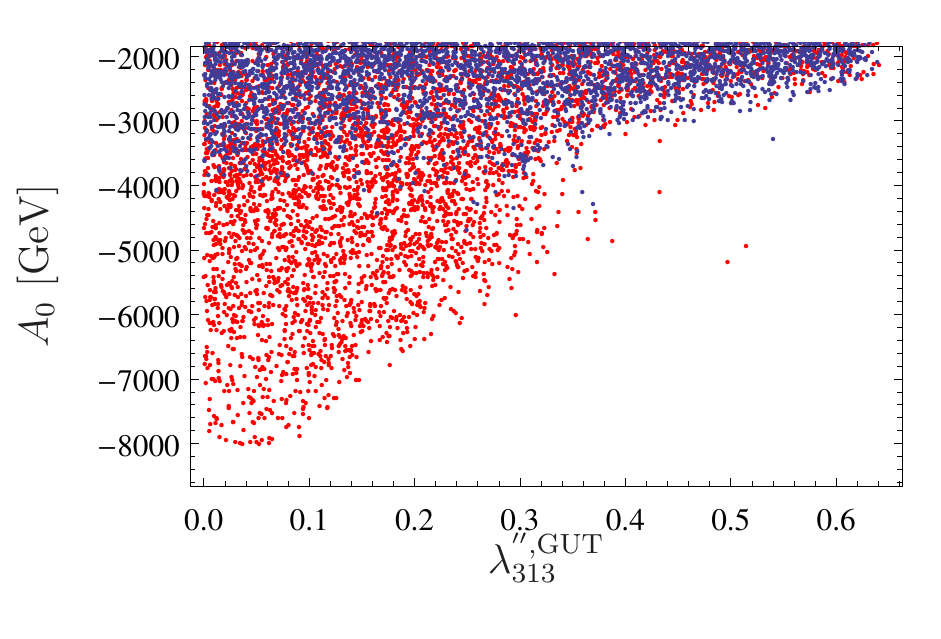} 
\caption{Allowed values for $A_0$ as function of $\lam^{'',{\scriptsize\mbox{GUT}}}_{313}$. The blue points 
have a stable EW vacuum while for the red points deeper CCB vacua exist. No cut is imposed on the light Higgs mass. 
}
\label{fig:A0}
\end{figure}

\begin{figure}[tb]
\centering
\includegraphics[width=0.49\linewidth]{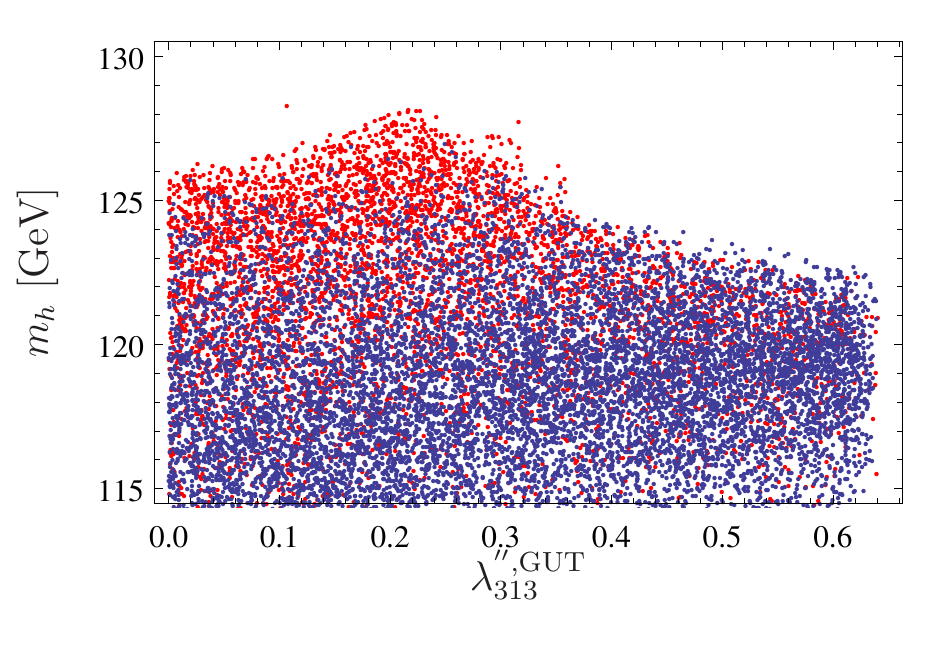} \hfill 
\includegraphics[width=0.49\linewidth]{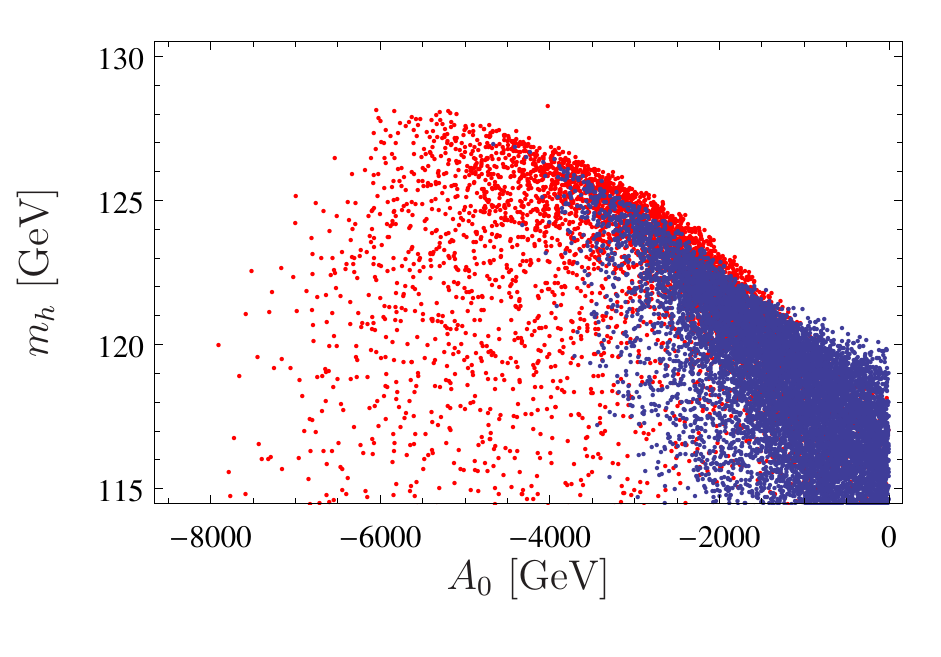} \\
\caption{Dependence of the light Higgs mass $m_h$ on $\lam^{'',{\scriptsize\mbox{GUT}}}_{313}$ (left) 
and $A_0$ (right). All points fulfil $m_{\tilde{t}_1} < 1$~TeV. 
The blue points have a stable EW vacuum while for the red points 
deeper CCB vacua exist. In the scan, no constraint has been imposed on 
the Higgs mass here. We do require $m_{\tilde t_1}< 1\,\mathrm{TeV}$}
\label{fig:mh}
\end{figure}

We see that the Higgs mass limit has a large impact on the preferred regions in the ($\lam^{''}_{31i}$, $m_ 
{\tilde{t}_1}$) plane: if no cut on the light Higgs mass is applied, the full planes shown in 
Fig.~\ref{fig:result313} are populated with (small empty grey) circles which have a stable EW vacuum. However, 
using $m_h \in [122,130]$~GeV makes it much more difficult to find viable points with  $\lam^{''}_{31i}>$
0.4, and stop masses below 1~TeV. This can be understood from Fig.~\ref{fig:A0}, where we show the 
correlation between $A_0$ and $\lam^{''}_{313}$: If $\lam^{''}_{313}$ increases, the upper limit of $|
A_0|$ allowed by a non--tachyonic spectrum decreases. For $\lam_{313}^{''} > 0.4$ a spectrum without 
tachyons requires $A_0 > -3000$~GeV. For larger values of $|A_0|$, the $T''_{\lam}$ contributions to the 
running of $m_{\tilde U}^2$ shown in Eq.~(\ref{eq:RGEmu2}) cause a negative soft SUSY breaking mass 
squared term for the right--handed stop. However, these values of $|A_0|$ are not sufficient to lift the light Higgs mass 
above the lower limit of 122~GeV, if the stop is too light. As a consequence, the maximal value for
 the light Higgs mass that we find decreases with increasing $\lam^{''}_{31i}$, because of the simultaneously 
 decreasing stop mass.   This behaviour can also be seen in Fig.~\ref{fig:mh}. 

For small or vanishing $\lam^{''}_{31i}$ couplings and a Higgs mass above 124~GeV the minimal stop mass 
with a stable EW vacuum is above 800~GeV. This is in agreement with our expectations based on 
Fig.~\ref{fig:CMSSM}. In contrast, we find for $\lam_{313}^{''} \sim 0.3$ points with stop masses below 
400~GeV which do not suffer from a deeper CCB vacuum. 

If we relax the bound on the heavy Higgs mass and use as the lower limit 122~GeV, we find in the 
$R$--parity conserving limit ($\lam''_{31i} \ll 0.1$) already points with stop masses below 350~GeV. However, 
for these small $\lam''_{31i}$ couplings a theoretical uncertainty of 4~GeV on the light Higgs mass might be 
overestimated. In addition, these points usually have a small value of $M_{1/2}$, as seen in 
Fig.~\ref{fig:CMSSM}. This implies a light gluino mass which would be in conflict with current mass limits 
\cite{Aad:2014wea}, for $\lam''_{31i} = 0$. Furthermore, in the case of non--zero $\lam''_{31i}$, constraints from 
LHC searches for three--jet resonances from gluinos apply~\cite{Chatrchyan:2012uxa,Chatrchyan:2013gia}. 
Thus it is questionable if these points should be considered at all. Nevertheless, also for this very conservative 
limit on the Higgs mass, one can find parameter points with even lighter stops if sizeable $R$pV couplings are 
present. In general, we find in our scans that $\lam_{313}^{''} \simeq 0.3$ turns out to be the optimal value to find 
parameter points with a light stop with $m_{\tilde{t}_1} \sim (220 - 250)$~GeV, a Higgs mass in agreement with 
the measurement and a stable EW vacuum. 

\begin{figure}[tb]
\centering
\includegraphics[width=0.49\linewidth]{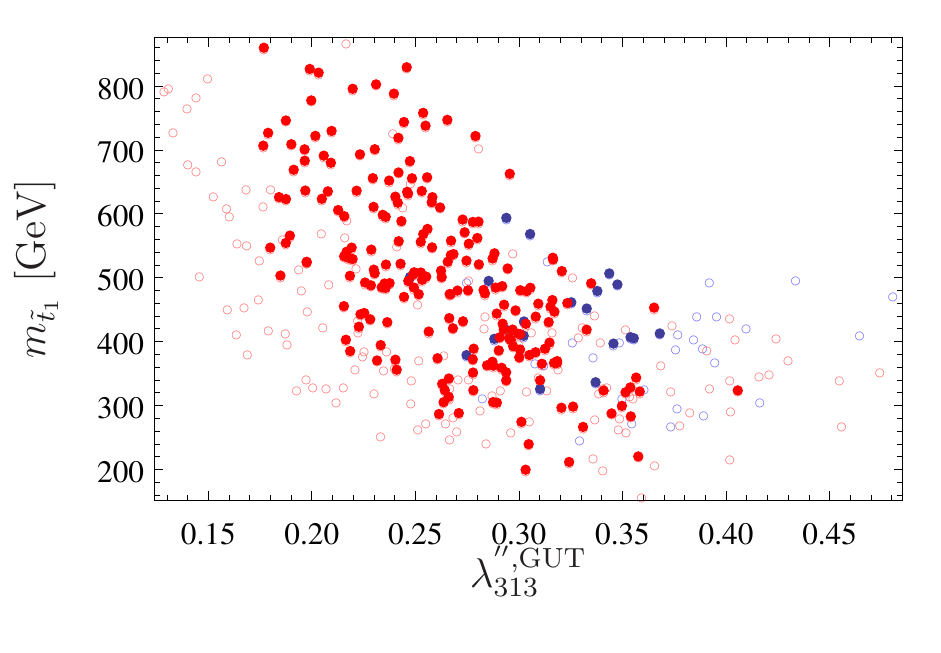} \hfill
\includegraphics[width=0.49\linewidth]{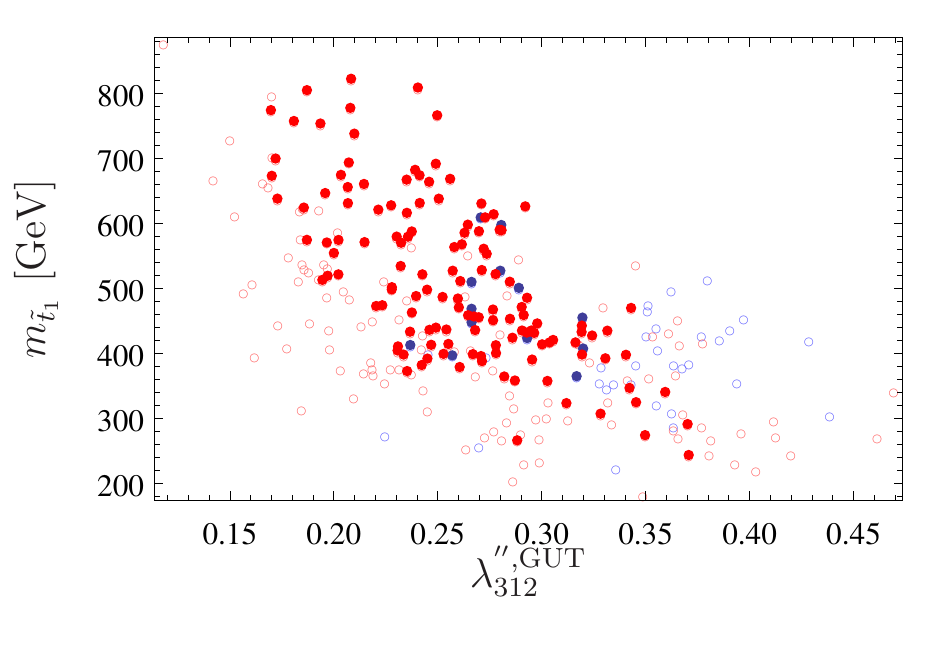} 
\caption{Vacuum stability in the $(\lam^{'',{\scriptsize\mbox{GUT}}}_{313},m_{\tilde{t}_1})$ 
plane (on the left) and $(\lam^{'',{\scriptsize\mbox{GUT}}}
_{312},m_{\tilde{t}_1})$ plane (on the right) showing only points with a stop lightest supersymmetric particle 
(LSP). The blue points have a stable EW vacuum while for the red points deeper CCB vacua exist. For the 
filled dots we required $m_h \in [124,128]$~GeV, while for the empty circles we only required $m_h \in 
[122,130]$~GeV.}
\label{fig:resultLSP}
\end{figure}

In Fig.~\ref{fig:resultLSP} we show the same planes as in Fig.~\ref{fig:result313}, but only for points where the 
lightest supersymmetric particle (LSP) is the stop. In principle, it is possible to have a stop LSP in the CMSSM 
without $R$pV operators \cite{Ellis:2014ipa}. However, these regions are usually very fine--tuned and need very 
large values of $M_{1/2}$ in order to raise the $\tilde\chi^0_1$ mass and to obtain a light Higgs mass in the experimentally preferred range. Therefore, we found no points with a stop LSP and a moderately small RpV coupling $\lam^{''}_{31i} < 0.14$ in our scan. In contrast, for larger values of the $R$pV 
couplings it is much easier to find a stop LSP. Also here we find that the minimal stop mass is obtained for 
$\lam^{''}_{31i} \simeq 0.3$. We emphasise that the points in Fig.~\ref{fig:result313}, which feature very light stop masses, are exactly the same points as in Fig.~\ref{fig:resultLSP}, which have a stop LSP. This is a non--trivial observation, because $M_{1/2}$ must be large to lift the $\tilde\chi^0_1$ mass above the $\tilde t_1$ mass. However, at the same time, a large value of $M_{1/2}$ also raises the $\tilde t_1$ mass via the RGEs.

\begin{figure}[tb]
\centering
\includegraphics[width=0.49\linewidth]{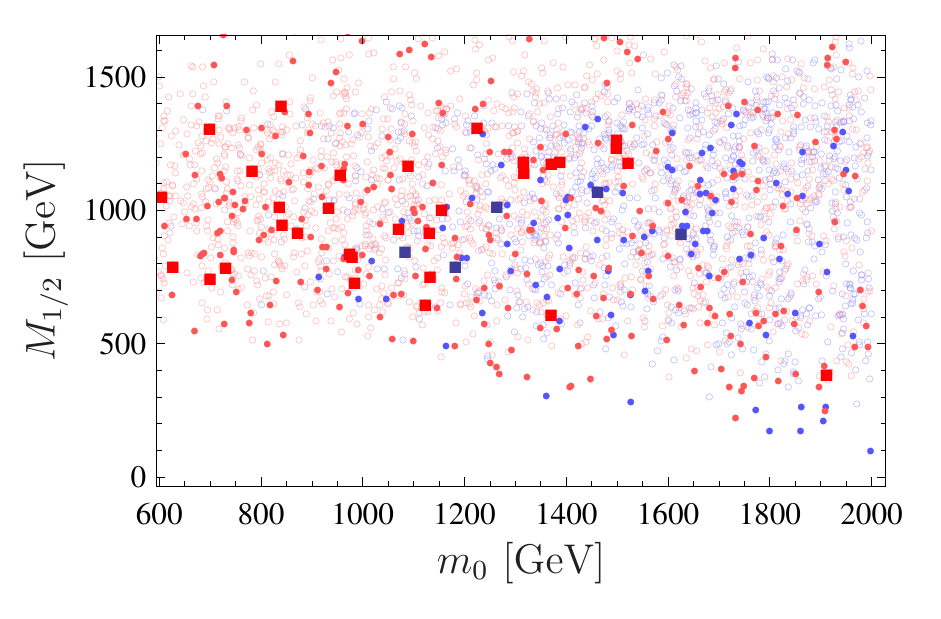} \hfill
\includegraphics[width=0.49\linewidth]{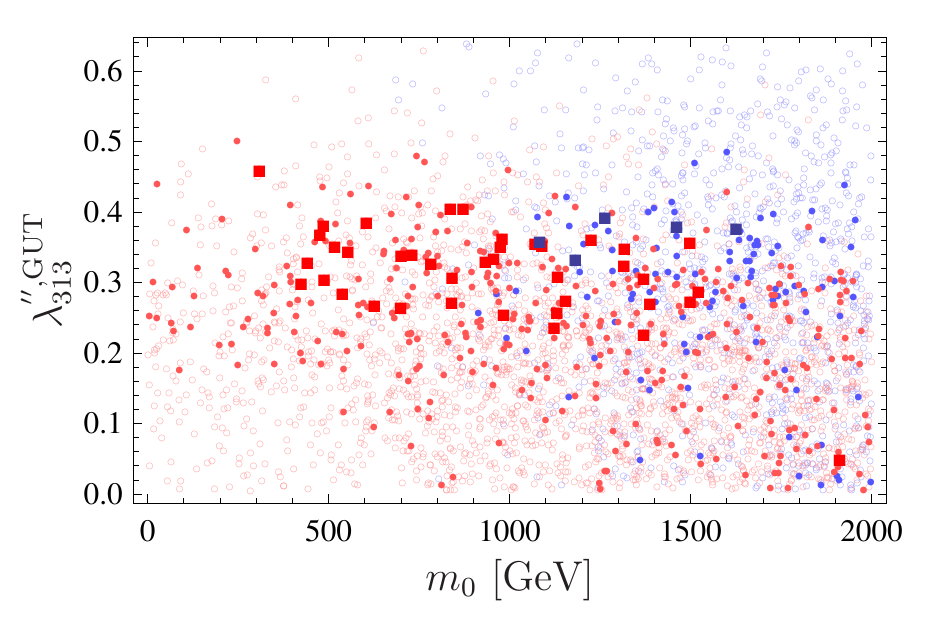} 
\caption{{The mass of the lightest stop in the 
$(m_0, M_{1/2})$ plane (left)  and $(m_0,\lam^{'',\scriptsize\mbox{GUT}}_{313})$ plane (right). 
Here, we required $m_h \in [122,130]$~GeV. The blue points have a stable  EW vacuum while for the red points 
deeper CCB vacua exist. The mass of the light stop is indicated by the plot marker: $m_{\tilde{t}_1} < 1$~TeV 
(open circles), $m_{\tilde{t}_1} < 0.5$~TeV (filled circles), $m_{\tilde{t}_1} < 0.3$~TeV (filled squares). }}
\label{fig:mStopInput}
\end{figure}

One might wonder how strong the bias from our parameter choice in Eqs.~(\ref{eq:ranges}) and (\ref{eq:ranges2})
is: It might be possible to find very light stops fulfilling all considered constraints for even larger values of 
$\lam^{''}_{{31i}}$, if the maximal value of $m_0$ or $M_{1/2}$ is increased. However, this is not the case because this 
would also increase the size of the radiative corrections to the light stop, as the squarks of the first and 
second generation also get heavier. To demonstrate that our points with very light stops are not on the edge of 
our parameter range, we show the correlation between the mass of the light stop and $m_0$, $M_{1/2}$ and $
\lam^{''}_{{31i}}$, respectively, in Fig.~\ref{fig:mStopInput}.  The red and blue filled squares denote a stop mass
$m_{\tilde{t}_1}<0.3$~TeV. These are not clustered at the edge of our allowed ranges. In fact, in the $R$-parity
violating coupling the low--mass stops are clustered around $\lam^{''}_{313}\sim0.3$.

We conclude with a brief comment on the collider aspects of the presented scenario. To this affect, we have selected four 
benchmark points (BP313A, BP313B, BP312A, BP312B) which show the main characteristics of the mass spectra in our preferred 
parameter regions with very light stops. We give in Table~\ref{main-bench} only the most important values but 
list the entire spectrum and input values in Appendix~\ref{app:benchmark}.

\begin{table}
\begin{center}
\begin{tabular}{|c|cc|cc|}
\hline 
  & BP313A & BP313B & BP312A & BP312B \\
\hline 
$m_{h_1}$ [GeV]              & 124.5  &  122.3 & 124.4  & 122.9  \\
$m_{\tilde{t}_1}$ [GeV]      & 325.8  &  247.9 & 364.6  & 222.9 \\
$m_{\tilde{t}_2}$ [GeV]      & 2473.3 &  1670.9& 2227.0 & 1719.6 \\
$m_{\tilde{b}_1}$ [GeV]      & 2015.1 &  1353.5& 2215.7 & 1703.7 \\
$m_{\tilde{d}_R}$ [GeV]      & 2075.4 &  1420.8& 1896.0 & 1437.3 \\
$m_{\tilde{s}_R}$ [GeV]      & 2792.8 &  1908.1& 1896.0 & 1437.3 \\
$m_{\tilde{q}}$ [GeV]      & $>$ 2800 &  $>$ 1900& $>$ 2500 & $>$ 1950 \\
$m_{\tilde{\tau}_1}$ [GeV]   & 1441.1 &  1139.3& 1446.7 & 976.1 \\
$m_{\tilde{\chi}^0_1}$ [GeV] & 568.2  &  334.9 & 480.5  & 373.3 \\
$m_{\tilde{\chi}^+_1}$ [GeV] & 1073.2 &  639.1 & 911.3  & 710.6 \\
$m_{\tilde{g}}$ [GeV]        & 2834.9 &  1794.0& 2461.0 & 1955.9 \\
\hline
\end{tabular}
\end{center}
\caption{Main features of our benchmark points. The full information is given in Appendix~\ref{app:benchmark}.
The benchmarks BP313 have $\lam^{''}_{313}\not=0$, while the benchmarks BP312 have $\lam^{''}_{312}\not=0$. $m_{\tilde q}$ refers to all not explicitly listed squark masses. $\tilde{t}_1$ is the LSP.}
\label{main-bench}
\end{table}

\medskip

In Table~\ref{main-bench}, we have summarised all squark masses not given explicitly by $m_{\tilde{q}}$. While the stop mass can 
be reduced due to the large $R$pV couplings, all other SUSY scalars are heavy and typically in the multi--TeV range. Also, the 
gluino is very heavy, and just the electroweakinos can have masses below 1~TeV. Thus for these benchmarks $\tilde{t}_1$ is the 
LSP. This mass hierarchy together with the large $R$pV couplings, yields prompt di--jet decays of the stop LSP, and
makes it difficult to look for the light stops at the LHC \cite{Chatrchyan:2013izb,Chatrchyan:2011ns,Chatrchyan:2013qha,Chatrchyan:2013gia,Aad:2011fq,ATLAS:2012pu,ATLAS:2012qjz,Evans:2012bf,Duggan:2013yna,Bai:2013xla}. 
Therefore, we leave the exploration of possible search strategies for such light stops together with large $R$pV couplings for future 
studies. 

\section{Conclusion}
\label{sec:conclusion}
We have discussed the possibility of light stops in a constrained version of the MSSM extended by large 
$R$--parity violating couplings $\lam^{''}_{31i},$ $i=2,3$. It has been shown that in this model it is possible 
to find parameter regions providing light stops with masses as low as 250~GeV which are consistent with 
the Higgs mass measurement, flavour observables and the stability of the electroweak vacuum. This is different 
from the CMSSM without $\bar{\bf U}\bar {\bf D}  \bar {\bf D}$ operators where  large stop mixing 
or heavy stops are needed to accommodate the Higgs mass. There the presence of light stops is highly 
constrained by the stability of the electroweak vacuum. Thus stop masses below 800~GeV can hardly be 
obtained in the $R$--parity conserving CMSSM. In the CMSSM with large $R$--parity violation, an interesting observation is that the lightest stop 
mass is usually found for $\lam''_{31i}\simeq 0.3$. In these scenarios the lightest stop is usually the LSP. We have shown 
that for this size of $R$pV couplings it is necessary to calculate the additional $R$pV one--loop corrections to the 
stop mass. 
These corrections can alter the prediction of the light stop mass by more than 
100~GeV compared to an incomplete one--loop calculation that takes into account only $R$--parity conserving interactions. 

\section*{Acknowledgements}
We thank Ben Allanach for his fast replies to our questions regarding {\tt SoftSUSY} and Ben O'Leary for many 
interesting discussions about vacuum stability. NC is supported by the Alexander von Humboldt foundation.
HD, FS and TS are supported by the BMBF PT DESY Verbundprojekt 
05H2013-THEORIE `Vergleich von LHC-Daten mit supersymmetrischen Modellen'.

\begin{appendix}
\section{Benchmark points}
\label{app:benchmark}

In Tab.~\ref{full-bench} we list the explicit parameters, the full sparticle mass spectrum and the predictions for the relevant flavour observables of our four benchmark scenarios. The $R$pV coupling is evaluated 
at the GUT scale and is about 0.3. This results in the lowest possible stop masses in our scan. The sparticle masses are all above 
1~TeV, except those of the lightest neutralino $\tilde\chi^0_1$ and the lightest stop $\tilde t_1$. The latter is the LSP for all four
benchmark points. Thus we expect the dominant stop decay to be to two jets: $\tilde t\to d d_i$, with possibly a $b$--jet for
$i=3$.

\begin{table}[!h]
\begin{center}
\begin{tabular}{|c|cc|cc|}
\hline 
  & BP313A & BP313B & BP312A & BP312B \\
\hline 
$m_0$ [GeV]                  & 1437.8 &  1182.0& 1466.8 & 1075.7 \\
$M_{1/2}$ [GeV]              & 1299.1 &  780.5 & 1104.4 & 867.6  \\
$\tan\beta$                  & 12.8   &   17.2 & 14.4   & 22.1   \\
sign($\mu$)                  &   +    &    +   &  +     &  +     \\
$A_0$ [GeV]                  &-3555.5 & -2152.8& -2972.3& -2347.6\\
\hline 
$\lam^{'',\scriptsize\mbox{GUT}}_{313}$              & 0.310  &  0.329 & 0      & 0     \\
$\lam^{'',\scriptsize\mbox{GUT}}_{312}$              & 0      &  0     & 0.317  & 0.335  \\
\hline 
$m_{h_1}$ [GeV]              & 124.5  &  122.3 & 124.4  & 122.9  \\
$m_{h_2}$ [GeV]              & 2523.7 &  1655.1& 2238.6 & 1571.6 \\
$m_{A}$ [GeV]                & 2554.0 &  1668.4& 2253.1 & 1613.3 \\
$m_{H^+}$ [GeV]              & 2523.7 &  1656.3& 2238.9 & 1573.4 \\
$m_{\tilde{t}_1}$ [GeV]      & 325.8  &  247.9 & 364.6  & 222.9 \\
$m_{\tilde{t}_2}$ [GeV]      & 2473.3 &  1670.9& 2227.0 & 1719.6 \\
$m_{\tilde{b}_1}$ [GeV]      & 2015.1 &  1353.5& 2215.7 & 1703.7 \\
$m_{\tilde{b}_2}$ [GeV]      & 2463.7 &  1658.4& 2469.4 & 1867.8 \\
$m_{\tilde{d}_L}$ [GeV]      & 2892.5 &  1961.3& 2609.9 & 2038.0 \\
$m_{\tilde{d}_R}$ [GeV]      & 2075.4 &  1420.8& 1896.0 & 1437.3 \\
$m_{\tilde{u}_L}$ [GeV]      & 2891.6 &  1959.9& 2608.9 & 2036.7 \\
$m_{\tilde{u}_R}$ [GeV]      & 2803.8 &  1914.1& 2539.2 & 1981.3 \\
$m_{\tilde{s}_R}$ [GeV]      & 2792.8 &  1908.1& 1896.0 & 1437.3 \\
$m_{\tilde{\tau}_1}$ [GeV]   & 1441.1 &  1139.3& 1446.7 & 976.1 \\
$m_{\tilde{\tau}_1}$ [GeV]   & 1640.5 &  1256.7& 1603.6 & 1159.7 \\
$m_{\tilde{l}_R}$ [GeV]      & 1512.7 &  1215.0& 1519.8 & 1120.1 \\
$m_{\tilde{l}_L}$ [GeV]      & 1670.4 &  1288.0& 1634.7 & 1218.2 \\
$m_{\tilde{\chi}^0_1}$ [GeV] & 568.2  &  334.9 & 480.5  & 373.3 \\
$m_{\tilde{\chi}^0_2}$ [GeV] & 1073.1 &  639.0 & 911.2  & 710.5 \\
$m_{\tilde{\chi}^0_3}$ [GeV] & 2019.9 &  1227.1& 1702.0 & 1334.3 \\
$m_{\tilde{\chi}^0_4}$ [GeV] & 2023.1 &  1231.8& 1705.6 & 1338.6 \\
$m_{\tilde{\chi}^+_1}$ [GeV] & 1073.2 &  639.1 & 911.3  & 710.6 \\
$m_{\tilde{\chi}^+_2}$ [GeV] & 2023.5 &  1232.6& 1706.2 & 1339.3 \\
$m_{\tilde{g}}$ [GeV]        & 2834.9 &  1794.0& 2461.0 & 1955.9 \\
\hline  
$R(b\to s \gamma)$           & 0.98   &  0.90  & 0.96   & 0.86 \\
$R(B\to\mu \nu)$             & 0.99   &  0.98  & 0.99   & 0.98 \\ 
$R(B_s\to \mu^+ \mu^-)$      & 1.14   &  1.18  & 1.13   & 1.25 \\
$R(B_d\to \mu^+ \mu^-)$      & 1.14   &  1.16  & 1.13   & 1.24 \\
$R(\Delta M_{B,s})$          & 1.01   &  1.02  & 1.01   & 1.02 \\
$R(\Delta M_{B,d})$          & 1.01   &  1.02  & 1.01   & 1.02 \\
$R(\epsilon_K)$              & 1.01   &  1.02  & 1.02   & 1.02 \\
\hline
\end{tabular}
\end{center}
\caption{Full sparticle mass spectrum and flavour observables predicted for our benchmark points. The benchmarks BP313 have $\lam^{''}_{313}\not=0$, while the benchmarks BP312 have $\lam^{''}_{313}\not=0$.}
\label{full-bench}
\end{table}

 \clearpage

\section{$\bar{\bf U}\bar {\bf D}
  \bar {\bf D}$ corrections to stop masses}
\label{app:stop}
\begin{figure}[h]
\includegraphics[width=\linewidth]{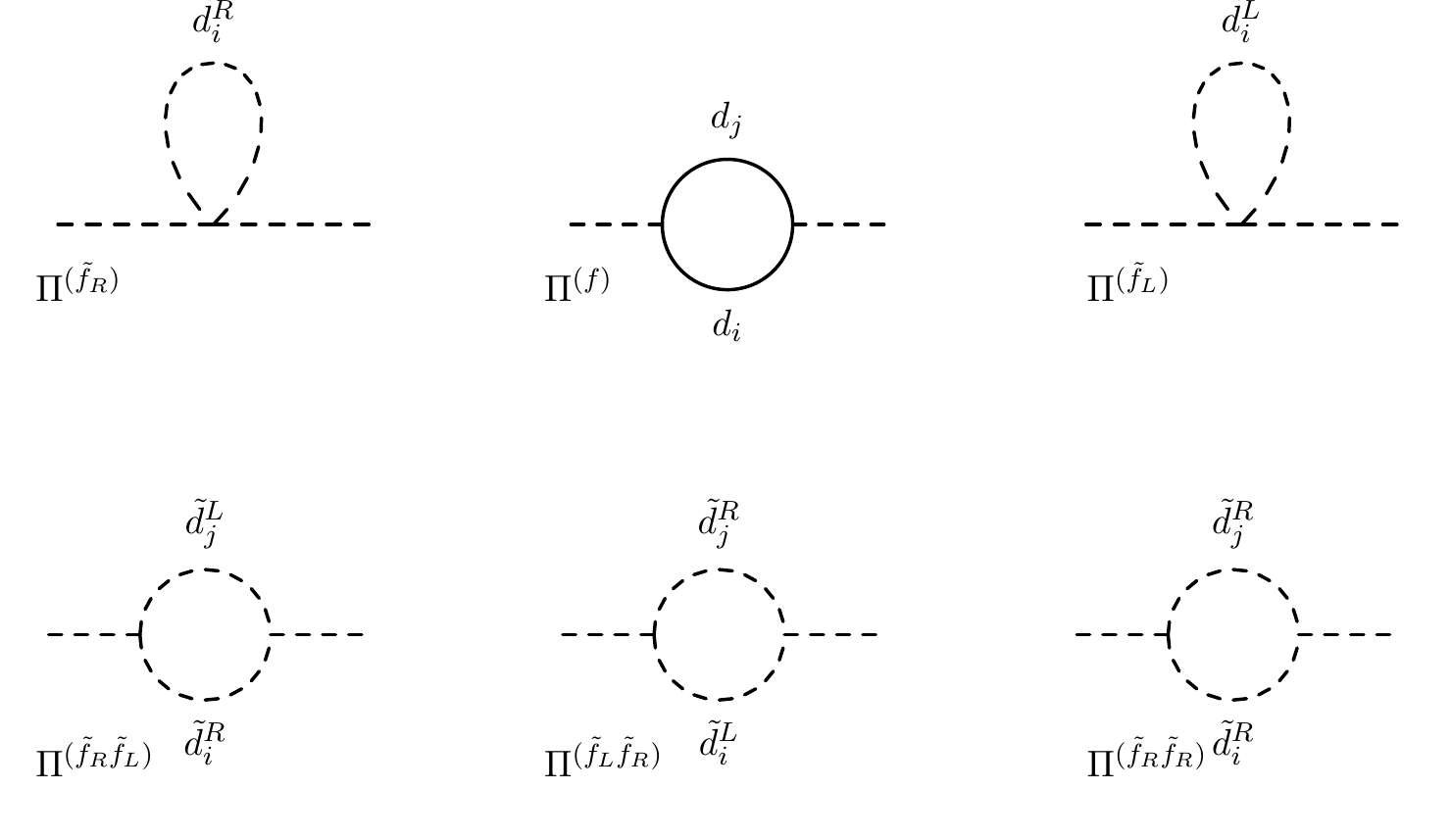}
\caption{One-loop correction to the stop mass due to down--like (s)quarks.}
\label{fig:OneLoopDiagrams}
\end{figure}

In the following we give a brief analytical estimate of the one--loop corrections to the stop masses in the 
presence of large $\lam''$ couplings.  The necessary Feynman diagrams are shown in  
Fig.~\ref{fig:OneLoopDiagrams}. This also defines our notation for the two--point functions $\Pi^{(\cdot)}$.

We start with the corrections which only involve superpotential couplings: since $\Pi^{\tilde f_L}$ 
(see Fig.~\ref{fig:OneLoopDiagrams}) has no contribution from baryon number violating operators, we do not 
consider it in the following. $\Pi^{\tilde f_R \tilde f_R}$ only has contributions proportional to the soft--terms 
$T''_\lam$, which will be discussed below. The amplitudes for the remaining diagrams can be expressed by
\begin{align}
16 \pi^2 \Pi^{ff} =& (|\Gamma^L(\tilde{t}_R,d_i,d_j)|^2 +|\Gamma^R(\tilde{t}_R,d_i,d_j)|^2)  G(p^2, m_{d_i}^2, m_{d_j}^2) \nonumber \\
& - 2 (\Gamma^L(\tilde{t}_R,d_i,d_j) \Gamma^R(\tilde{t}_R,d_i,d_j)^* \nonumber \\
& + \Gamma^L(\tilde{t}_R,d_i,d_j)^* \Gamma^R(\tilde{t}_R,d_i,d_j))  m_{d_i} m_{d_j} B_0(p^2,m_{d_i}^2, m_{d_j}^2)\,, \\
16 \pi^2 \Pi^{\tilde{f} \tilde{f}} =& |\Gamma(\tilde{t}_R, \tilde{d}^R_i, \tilde{d}^L_j)|^2 B_0(p^2,m_{\tilde{d}^R_i}^2,m_{\tilde{d}^L_j}^2) + (i\leftrightarrow j)\,, \\
16 \pi^2 \Pi^{\tilde{f}} =& - \Gamma(\tilde{t}_R, \tilde{t}_R, \tilde{d}^R_i, \tilde{d}^R_i) A_0(m_{\tilde{d}^R_i}^2) + (i\leftrightarrow j)\,.
\end{align}
with $\Pi^{\tilde{f} \tilde{f}} \equiv \Pi^{\tilde{f}_L \tilde{f}_R} + \Pi^{\tilde{f}_R \tilde{f}_L}$. 
Here, we have introduced
\begin{equation}
G(p^2, m_1^2, m_2^2) \equiv - A_0(m_1^2) - A_0(m_2^2) + (p^2 - m_1^2 - m_2^2) B_0(p^2,m_1^2,m_2^2)  \, .
\end{equation}
$A_0$ and $B_0$ are the standard Passarino-Veltman integrals \cite{Passarino:1978jh}. 
The $\Gamma$'s represent the involved vertices. 
These are given in the limit 
of diagonal Yukawa couplings by:
\begin{enumerate}
 \item (Chiral) stop--quark--quark vertex:
 \begin{eqnarray*}
  \Gamma^L(\tilde{t}_R,d_i,d_j) &\equiv& \lam^{''}_{3ij} \,,\\
  \Gamma^R(\tilde{t}_R,d_i,d_j) &=& 0  \,.
 \end{eqnarray*}
 \item Stop--squark--squark vertex: 
 \begin{equation}
  \Gamma(\tilde{t}_R, \tilde{d}^R_i, \tilde{d}^L_j) \equiv \lam^{''}_{3ij} Y_d^{jj} \langle H_d \rangle = m_{d_j} 
  \lam^{''}_{3ij}\,.
 \end{equation}
 \item Four squark vertex:
 \begin{equation}
  \Gamma(\tilde{t}_R, \tilde{t}_R, \tilde{d}^R_i, \tilde{d}^R_j) \equiv - \lam^{''}_{3ik} \lam^{''}_{3jk}\,.
 \end{equation}
\end{enumerate}

One can check easily that in the limit of unbroken SUSY, $m_{\tilde{d}^R_i} = m_{\tilde{d}^L_i} = m_{d_i}$,
the sum of all diagrams vanishes exactly
\begin{equation}
\Pi^{ff} + \Pi^{\tilde{f}\tilde{f}}  + \Pi^{\tilde{f}} = 0 \, .
\end{equation}
If we assume for simplicity that all SUSY masses are degenerate ($m_{\tilde{d}^R_i} = m_{\tilde{d}^L_i} = M_{SUSY},$ $\forall i$) 
and take the limit $p^2 \to 0$, $M_{SUSY} 
\gg m_{d_i}$ we obtain a very simple expression for the sum of all diagrams
\begin{equation}
\Pi^{ff} + \Pi^{\tilde{f}\tilde{f}}  + \Pi^{\tilde{f}} = \frac{1}{8\pi^2}  |\lam^{''}_{3ij}|^2 M_{SUSY}^2 \, \equiv \Pi^{\lam^{''}} .
\end{equation}
Here, we have used 
\begin{align}
B_0(0,m^2,m^2) = &  \frac{A_0(m^2)}{m^2} - 1 \,,\\
A_0(m^2) = & m^2 - m^2 \log\frac{m^2}{Q^2}\,,
\end{align}
and set as the renormalisation scale $Q=M_{SUSY}$.

As mentioned above there is also another one--loop contribution due to the trilinear soft-breaking terms:
\begin{equation}
16 \pi^2 \Pi^{\tilde{f}_R \tilde{f}_R} = |T^{''}_{3ij}|^2 B_0(p^2,m_{\tilde{d}^R_i}^2,m_{\tilde{d}^R_j}^2) \, . 
\end{equation}
However, this contribution vanishes exactly in the limit of degenerate down--like squark masses
$m_{\tilde{d}^R_i} = m_{\tilde{d}^R_j} = Q = M_{SUSY}$. 
Hence, it can only play a role in the case of a large mass splitting between the squarks in the loop. To see this, we can use $m_{\tilde{d}^R_i} =  Q = M_{SUSY}$ together with $m_{\tilde{d}^R_j} = M_{SUSY} + \delta$ and obtain 
\begin{equation}
\Pi^{\tilde{f}_R \tilde{f}_R} = - \frac{\delta^2}{16 \pi^2 M_{SUSY}^2} |T^{''}_{3ij}|^2 \,.
\end{equation}
Here, we have made use of 
\begin{equation}
B_0(0,m^2_1, m^2_2)= - \log\frac{m^2_2}{Q^2} + \frac{1}{m^2_2 - m^2_1}\left(m^2_2 - m^2_1 + m^2_1 \log\frac{m^2_1}{m^2_2}\right)\,.
\end{equation}

We can now use the derived expressions for the one--loop self-energies to calculate the stop mass at 
one--loop. If we neglect flavour mixing in the squark sector, the one--loop stop masses are the eigenvalues of the 
one--loop corrected stop mass matrix $m^{2,(1L)}_{\tilde t}$ given by
\begin{equation}
m^{2,(1L)}_{\tilde t} = m^{2,T}_{\tilde t} + \delta m^{2,MSSM} + (\Pi^{\lam^{''}} + \Pi^{\tilde{f}_R \tilde{f}_R}) \left(\begin{array}{cc} 0 & 0 \\ 0 & 1\end{array} \right)\,.
\end{equation}
Here, $m^{2,T}_{\tilde t}$ is the stop mass matrix at tree--level, 
\begin{eqnarray}
m^{2,T}_{\tilde t} &=& \left(\begin{array}{cc} 
m_{\tilde t_L}^2  -\frac{1}{24} \Big(g_{1}^{2}-3 g_{2}^{2}\Big) \Big(v_{d}^{2}- v_{u}^{2}  \Big)  +   \frac{v_{u}^{2}}{2} |Y_{t}|^2 &
\frac{1}{\sqrt{2}} \Big(v_u T_{t}^{*}- v_d \mu Y_{t}^{*} \Big) \\ 
\frac{1}{\sqrt{2}}\Big( v_u T_{t}- v_d Y_{t} \mu^* \Big) & 
 m_{\tilde t_R}^2  + \frac{ v_{u}^{2}}{2} |Y_{t}|^2 + \frac{1}{6} g_{1}^{2}  \Big(v_{d}^{2}- v_{u}^{2}\Big) 
                           \end{array}\right)\,, \nonumber \\[2mm]
\end{eqnarray}
and $\delta m^{2,MSSM}$ is 
the matrix for the well known corrections which do not involve $R$-parity violating couplings, see e.g. Ref.~\cite{Pierce:1996zz}.

\section{Minimising the scalar potential of the MSSM with $\sfb U \sfb D\sfb D$  operators}
\label{app:ScalarPotential}
We discuss in the following the scalar potential in the MSSM in the presence of $\lam^{''}_{3ij}$ couplings and 
vevs for stops, staus, as well as the down--type squarks at tree--level. The checks performed by \Vevacious include also
the one--loop corrections to the effective potential. However, these expressions are not shown here because of 
their length. To simplify the expressions we assume here that the Yukawa couplings and the soft-breaking 
parameters in the $R$-parity conserving sector are diagonal:
\begin{eqnarray*}
&Y_{\tau} = Y_{e,33},\hspace{1cm} Y_b = Y_{d,33},\hspace{1cm}Y_t = Y_{u,33}\,,&\\
&T_{\tau} = T_{e,33},\hspace{1cm} T_b = T_{d,33},\hspace{1cm}T_t = T_{u,33}\,,&\\
&m^2_{\tilde t_L} = m^2_{\tilde Q,33}, \hspace{1cm} m^2_{\tilde t_R} = m^2_{\tilde U,33}\,,&\\
&m^2_{\tilde \tau_L} = m^2_{\tilde L,33}, \hspace{1cm} m^2_{\tilde \tau_R} = m^2_{\tilde E,33}\,,&\\
&m^2_{\tilde q_i} = m^2_{\tilde Q,ii}, \hspace{1cm} m^2_{\tilde d_i} = m^2_{\tilde D,33}\,.&
\end{eqnarray*}

The full expression in the limit of diagonal Yukawa and $R$-parity conserving soft terms read
{\allowdisplaybreaks 
\begin{eqnarray}
V^{\text{tree}}_{H_{d}, H_{u}} &=& 
 \frac{1}{32}
 \left( g_{1}^{2} ( \vd{2} - \vu{2})^{2}
                    + g_{2}^{2} ( \vd{2} - \vu{2}  )^{2} \right)
 \nonumber \\ 
 & & - B_{\mu} \vd{} \vu{}
     + \frac{1}{2} \left( | \mu |^{2} ( \vd{2} + \vu{2} )
     + m_{H_{d}}^{2} \vd{2} + m_{H_{u}}^{2} \vu{2}
      \right)\,, \\
V^{\text{tree}}_{H_{d}, H_{u}, \sL, \sE}
 & = & \frac{1}{32}
 \left( g_{1}^{2} ( \vl{2} - 2 \ve{2} )^{2}
                    + g_{2}^{2} ( \vl{2} )^{2} \right)
 \nonumber \\
 & & + \frac{1}{4} \left( Y_{\tau}^{2} ( \vd{2} \vl{2}
     + \vd{2} \ve{2} + \vl{2} \ve{2} \right)
     + \frac{1}{ \sqrt{2} } \vl{} \ve{} \left(T_{\tau} \vd{}
     -  Y_{\tau}  \mu \vu{} \right)
 \nonumber \\ 
 & & + \frac{1}{2} \left(m_{\sL}^{2} \vl{2} + m_{\sE}^{2} \ve{2} \right)\,,
\label{eq:tree_Higgs_and_stau_potential} \\
V^{\text{tree}}_{H_{d}, H_{u}, \sTL, \sTR} & = & \frac{1}{288} \Big[3 \Big(4 \big(6 \big(\vTL{2} \left(2 m_{\tilde{t}_L}^2+Y_t^2 (\vTR{2}+\vu{2})\right)+\vTR{2}
   \left(2 m_{\tilde{t}_R}^2+\vu{2} Y_t^2 \right) \nonumber \\
    && +2 \sqrt{2} \vTL{} \vTR{} \left(\vu{} T_t-\vd{} Y_t \mu
   \right)\big)+g_3^2 (\vTL{2}-\vTR{2})^2\big)+ \nonumber \\
    && 3 g_2^2 \left(-2 \vu{2} \vTL{2}+2 \vd{2} \vTL{2}+\vTL{4}\right)\Big)+g_1^2 \big(\left(\vTL{2}-4
   \vTR{2}\right) (\vTL{2}-4 \vTR{2}+6 \vu{2}) \nonumber \\
    && -6 \vd{2} (\vTL{2}-4 \vTR{2})\big)\Big]\,, \\
V^{\text{tree}}_{H_{d}, H_{u}, \sDLi, \sDRi} & = & \frac{1}{288} \Big[3 \Big(4 \big(6 \big(\vDRi{2} \left(2 m_{\tilde{d}_i}^2+Y_{d_i}^2 (\vDLi{2}+\vd{2})\right)+\vDLi{2}
   \left(2 m_{\tilde{q}_i}^2+\vd{2} Y_{d_i}^2 \right) \nonumber \\
    && +2 \sqrt{2} \vDLi{} \vDRi{} \left(\vd{} T_{d_i}-\vu{} Y_{d_i} \mu
   \right)\big)+g_3^2 (\vDLi{2}-\vDRi{2})^2\big) \nonumber \\
   && +3g_2^2 \left(\vDLi{4}+2 \vDLi{2}
   (\vu{2}-\vd{2})\right)\Big)+g_1^2 \big(6 \vu{2} (\vDLi{2}+2
   \vDRi{2}) \nonumber \\
   && +(\vDLi{2}+2 \vDRi{2}) (\vDLi{2}-6 \vd{2}+2 \vDRi{2})\big)\Big]\,, \\
V^{\text{tree}}_{H_{d}, H_{u}, \sDLj, \sDRj} & = & V^{\text{tree}}_{H_{d}, H_{u}, \sDLi, \sDRi} \,\, | \,\, (i \to j)\,, \\
V^{\text{tree}}_{\sTL, \sTR,\sDLi, \sDRi} & = & \frac{1}{144} \Big[g_1^2 (\vDLi{2}+2 \vDRi{2}) (\vTL{2}-4 \vTR{2})-9 g_2^2 \vDLi{2}
   \vTL{2} \nonumber \\
   && -6 g_3^2 (\vDLi{2}-\vDRi{2})  (\vTL{2}-\vTR{2})\Big]\,, \\
V^{\text{tree}}_{\sTL, \sTR,\sDLj, \sDRj} & = & V^{\text{tree}}_{\sTL, \sTR,\sDLi, \sDRi}   \,\, | \,\, (i \to j) \,,\\ 
V^{\text{tree}}_{\sDLi, \sDRi,\sDLj, \sDRj} & = & \frac{1}{144} \Big[g_1^2 (\vDLj{2}+2 \vDRj{2}) (\vDLi{2}+2 \vDRi{2})+9g_2^2 \vDLj{2}
   \vDLi{2} \nonumber \\
   && -6g_3^2 (\vDLj{2}-\vDRj{2}) (\vDLi{2}-\vDRi{2}) +72 \vDLj{} \vDRj{}
   \vDLi{} \vDRi{} Y_{d_i} Y_{d_j}\Big]\,, \\
V^{\text{tree}}_{\sTR,\sTR, \sDRi, \sDRj} &=&  \frac{1}{4} \Big[\lam_{3ij}^{'',2} \left(\vDRj{2} (\vDRi{2}+\vTR{2})+\vDRi{2} \vTR{2}\right) \nonumber \\
&& -2 \lam_{3ij}^{''} \left(\vDLj{}
   \vd{} \vDRi{} \vTR{} Y_{d_j}+\vDRj{} \vDLi{} \vd{} \vTR{} Y_{d_i}+\vDRj{} \vDRi{} \vTL{} \vu{}
   Y_t \right) \nonumber \\
   && -2 \sqrt{2} T_{\lam_{3ij}^{''}} \vDRj{} \vDRi{} \vTR{}\Big]\,.
\end{eqnarray}
}
The $D$-term contributions are minimised for 
\begin{eqnarray}
& \vDLi{} = \vDRi{}, \hspace{1cm} \vDLj{} = \vDRj{}, \hspace{1cm} \vTL{}=\vTR{}, \hspace{1cm} v_u = v_d & \,.
\end{eqnarray}
In addition, 
for $j=3$ we neglect the terms involving $Y_{d_i}$ and $T_{d_i}$ which correspond 
to first or second generation Yukawas, respectively, trilinear terms. 
In this limit all terms involving down--squark vevs read
\begin{eqnarray}
V^{\text{tree}}_{\vDRi{}, \vDRj{}} = && \vDRi{2}\left(\frac{1}{2} (m_{\tilde{d}_i}^2+ m_{\tilde{q}_i}^2)+\frac{\lam_{3ij}^{'',2} \vTR{2}}{4}\right) \nonumber \\
&& +\vDRj{2}\left(\frac{1}{2} (m_{\tilde{d}_j}^2+ m_{\tilde{q}_j}^2)+\frac{\vd{} T_{d_j}}{\sqrt{2}}+\frac{\lam_{3ij}^{'',2} \vTR{2}}{4}+\frac{1}{2}
   \vd{2} Y_{d_j}^2-\frac{\vd{} Y_{d_j} \mu }{\sqrt{2}}  + \frac{\vDRj{2}}{4} Y_{d_j}^2 \right) \nonumber \\
&&+\frac{\vDRj{} \vDRi{}}{4} \left(\lam_{3ij}^{'',2} \vDRj{} \vDRi{} -2 \lam_{3ij}^{''} \vd{} \vTR{}
   (Y_{d_j}+Y_t)-2 \sqrt{2}  T_{\lam_{3ij}^{''}} \vTR{}\right) \nonumber \\
   && + \frac{g_1^2+g_2^2}{32} \left(\left(\vDRj{2}+\vDRi{2}\right) \left(\vDRj{2}+\vDRi{2}-2 \vTR{2}\right)\right)\,. 
\end{eqnarray}
The first line on the right hand side is always positive. Especially since the $\beta$-function of $m_{\tilde{q}_i}^2$ has no terms 
proportional to $\lam^{''}$ or to a third generation Yukawa coupling at one--loop it is positive and usually much larger than the other 
soft-parameter involved. This makes it rather unlikely that $\vDRi{} \neq 0$ are preferred at the minimum of the potential. However, in 
the limit $\vDRi{}\to 0$ the entire first and third line vanish. The form of the potential then has a similar form to  the 
potential when considering only stop and Higgs vevs. However, usually  $m_{\tilde{q}_i}^2 > m_{\tilde{t}_L}^2$ and 
$m_{\tilde{d}_i}^2 > m_{\tilde{t}_R}^2$ holds. This makes it  unlikely that the down--squarks gain a vev before the stops 
do. Thus, one can expect that the check for the additional down--squark vevs put only weak constraints in  addition for points with 
very large $\tan\beta$. 
While the discussion has been so far rather hand-waving the general statement has been confirmed in our numerical studies: Only 5\% of the points with stop masses below 1~TeV 
which pass the stability check including only stop vevs fail the additional test that includes sbottom vevs. Including sdown and sstrange vevs 
does not put any additional constraint.  For comparison: About 1/3 of the entire points in the scan fail the check 
for a stable vacuum when checking for stop and stau vevs. In the interesting parameter range of stop masses below 1~TeV even 2/3 are ruled out. 

\end{appendix}

\bibliographystyle{JHEP}
\bibliography{main}
\end{document}